\documentclass[onecolumn,showpacs,nofootinbib]{revtex4}

\usepackage{amsmath}
\usepackage{bbold}
\usepackage{array}
\usepackage[utf8]{inputenc}
\usepackage{graphicx} 
\usepackage{pstricks}
 \newcommand{\argtanh}{\mbox{\textsf{argth}}\,}
 \newcommand{\argsinh}{\mbox{\textsf{argsh}}\,}

\begin{document}

\title{ Nonintersecting Brownian Interfaces and Wishart Random Matrices}
\author{C\'eline Nadal and Satya N. Majumdar} 
\affiliation{Laboratoire de Physique Th\'{e}orique et Mod\`{e}les
Statistiques (UMR 8626 du CNRS), Universit\'e Paris-Sud, B\^atiment 100
91405 Orsay Cedex, France}




\begin{abstract}

We study a system of $N$ nonintersecting $(1+1)$-dimensional
fluctuating elastic interfaces (`vicious bridges') at thermal equilibrium,
each subject to periodic boundary condition in the longitudinal direction
and in presence of a substrate that induces 
an external confining potential for each interface.
We show that, in the limit of a large system
 and with an appropriate choice of the external confining potential,
the joint distribution of the heights of the $N$ nonintersecting interfaces
at a fixed point on the substrate 
can be mapped to the joint distribution of the eigenvalues of a Wishart matrix
of size $N$ with complex entries (Dyson index $\beta=2$), thus providing a
physical realization of the Wishart matrix. Exploiting this analogy
to random matrix, we calculate analytically (i) the average density
of states of the interfaces (ii) the height distribution of the
uppermost and lowermost interfaces (extrema) and (iii) the asymptotic (large $N$) distribution
of the center of mass of the interfaces. In the last case, we show
that the probability density of the center of mass has an
essential singularity around its peak which is shown to be a direct consequence
of a phase transition in an associated Coulomb gas problem.  
\end{abstract}

\maketitle

\section{Introduction}

The system of $N$ nonintersecting elastic lines was first studied by de Gennes~\cite{gennes} 
as a simple model of a fibrous structure made of $(1+1)$-dimensional
nonintersecting  flexible chains in thermal
equilibrium, under a unidirectional stretching force. These elastic lines
can also be viewed as the trajectories in time of $N$ nonintersecting
Brownian motions, a system studied in great detail by Fisher and co-workers~\cite{fisher-huse,fisher}
in the context of commensurate-incommensurate (C-IC) phase transitions. In this context
the nonintersecting lines are the domain walls between different
commensurate surface phases adsorbed on a crystalline substrate. 
The `nonintersection' constraint led Fisher to call this a problem
of `vicious' random walkers who do not meet (or kill
each other when they meet). Since then, the vicious walkers
model has had many physical applications, e.g., in wetting and melting~\cite{fisher-huse,fisher},
as a simple model of polymer network~\cite{Essam}, in the structure of
vicinal surfaces of crystals consisting of terraces divided by steps~\cite{Einstein,Richards}
and also in the context of stochastic growth models~\cite{Ferrari}. 

Depending on the underlying physical system being modelled by 
 $N$ vicious walkers, one can
pose and study a variety of statistical questions. For example, Huse and Fisher~\cite{fisher-huse} studied the so called 
`reunion' probability, i.e., the
probability $P(t)$ that $N$ vicious walkers starting at the same position 
in space reunite exactly after time $t$ at their same initial position but
without crossing each other in the time interval $[0,t]$ and showed
that it decays as a power law $P(t)\sim t^{-N^2/2}$ for large $t$.
Another pertinent issue is: given that the $N$ walkers have reunited
for the first time at time $t$, what can one say about the statistics
of the transverse fluctuations of the positions of the walkers
at any intermediate time $0\le \tau\le t$ (see Fig. 1)?
Such configurations where $N$ walkers emerge from a fixed point
in space and reunite at the same point after a fixed time $t$ are
called `watermelons' as their structure resembles
 that of a watermelon (see Fig. 1).
Such a watermelon configuration also describes the structure of the `droplet'
or the elementary topological excitation (vortex-antivortex pair) 
on the commensurate (ordered) side
of a C-IC phase transition~\cite{fisher-huse,fisher} with the longitudinal distance
between the pairs being $t$. The statistical (thermal) 
fluctuations
of the transverse sizes of such watermelon defects play an important role
near the phase transition. This initiated a study of the tranverse
fluctuations of the nonintersecting lines in the watermelon geometry
with fixed longitudinal distance $t$. In the random walk/probability language,
 this
means studying the transverse fluctuations of the trajectories of
$N$ walkers conditioned on the fact that they started and reunited
at the same point in space after a fixed time $t$ without crossing 
each other in between.
Another similar interesting geometrical configuration is
 a `watermelon with a wall', i.e,
$N$ nonintersecting walkers starting and reuniting at the same point in space
(say the origin) after a fixed time $t$, but staying positive in $[0,t]$ (see Fig. 1).
\begin{figure}
\includegraphics[width=\linewidth]{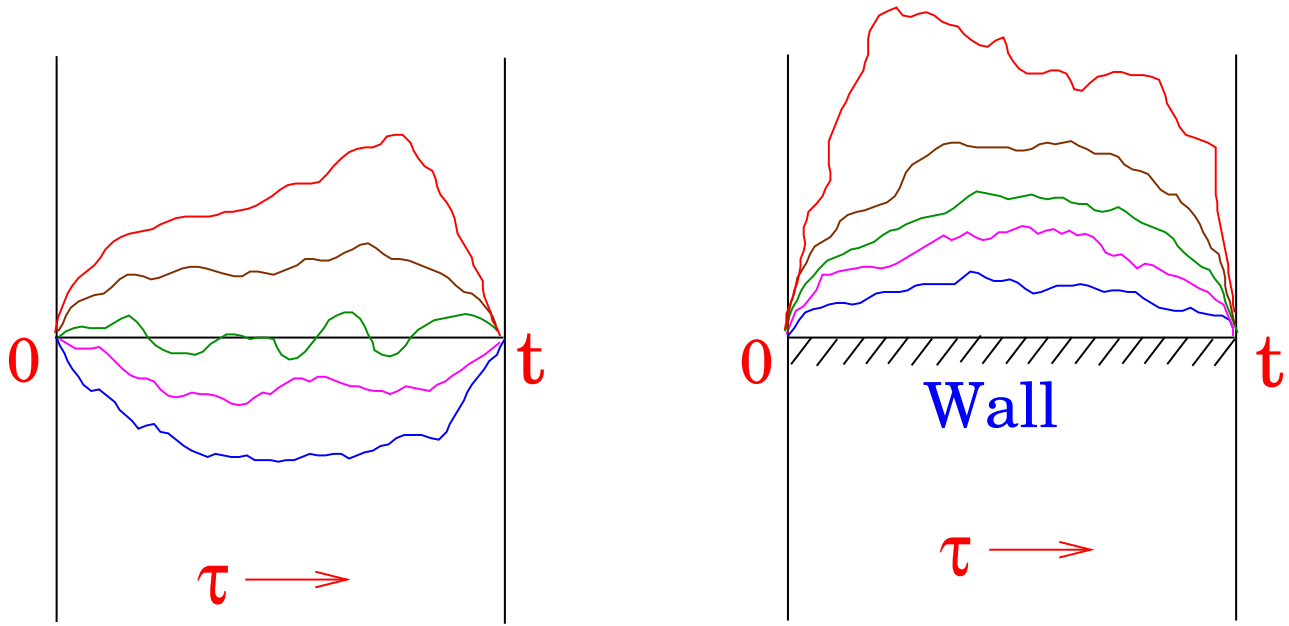}
\caption{Typical watermelon configurations without a wall (left) and with a wall (right)
at the origin for $N=5$ nonintersecting Brownian motions over the time interval $0\le 
\tau\le t$.}
\end{figure}

Recently, the transverse fluctuations of $N$ nonintersecting lines 
have been studied extensively
in watermelon
geometry over $[0,t]$ both with and without a wall and 
important connections to random matrix theory have been 
discovered~\cite{Ferrari,JohanssonRMT,Katori3,Katori2,Katori1, 
T-W,Schehr,N Kobayashi}. For example, the joint distribution of the positions of all the
walkers at a fixed time $0\le \tau \le t$ for watermelons without a wall 
was shown to be identical (after appropriate rescaling) to the joint distribution of eigenvalues
of a Gaussian random matrix belonging to the unitary (GUE) ensemble~\cite{JohanssonRMT,Katori1}.
On the other hand, the joint distribution of the positions at a fixed time $0\le \tau\le t$ 
for watermelons with
a hard wall at the origin was computed recently~\cite{Schehr} and was shown 
to be identical (after an appropriate change of variable) to the joint 
distribution of eigenvalues of 
a random matrix drawn from the Wishart (or Laguerre) ensemble at a special value of
its parameters, which also corresponds to the chiral Gaussian unitary ensemble
of random matrices \cite{Katori2}. 

It is useful at this point to recollect the definition of a Wishart matrix.
A Wishart matrix $W$ is an $(N\times N)$ square matrix of the product
form $W =X^{\dagger} X$ where $X$ is a $(M\times N)$ rectangular
matrix with real or complex entries and $X^{\dagger}$ is its Hermitian conjugate.
If the entries $X_{ij}$ represent some data, 
e.g., $X_{ij}$ may indicate the price of the $j$-th commodity on the $i$-th day, 
then $W$ is just the (unnormalized) covariance matrix that provides
informations about the correlations between prices of different commodities.
If $X$ is a Gaussian random matrix, $P(X)\propto \exp\left[-\frac{\beta}{2} {\rm 
Tr}(X^{\dagger} X)\right]$ where the Dyson index $\beta=1,2$ corresponds respectively to 
real and complex matrices,  
then the random covariance matrix $W$ belongs to the Wishart ensemble named after Wishart
who introduced them in the context of multivariate statistical data analysis~\cite{Wishart}.
Since then the Wishart matrix has found numerous applications.
Wishart matrices  play an important role in data compression techniques
such as  the ``Principal Components Analysis'' (PCA). 
PCA applications include image processing~\cite{Wilks,Fukunaga,Smith},
biological microarrays~\cite{arrays1,arrays2},  population
genetics~\cite{Cavalli,Patterson,genetics},
finance~\cite{BP,Burda},
meteorology
and oceanography~\cite{Preisendorfer}.
The spectral properties of the Wishart matrices have been studied extensively and it is 
known~\cite{James} that for $M\ge N$, all $N$ positive eigenvalues of $W$ are distributed
via the joint probability density function (pdf)
\begin{equation}
\label{eq:Wishart0}
P_N(\lambda_1,...,\lambda_N)= K_N\, e^{- \frac{\beta}{2}\,  \sum_k \lambda_k }
\prod_{k=1}^N \lambda_k^{\frac{\beta}{2}(M-N+1)-1}\,
\prod_{i<j} |\lambda_i-\lambda_j|^{\beta} 
\end{equation}
where $K_N$ is a normalization constant and the Dyson index
$\beta=1,2$
(respectively for real and complex $X$).
In the ``Anti-Wishart'' case, that is when $M<N$,
$W$ has $M$ positive eigenvalues ( and $N-M$ eigenvalues
that are exactly zero ) and their
joint probability distribution is simply obtained by
exchanging $M$ and $N$
in the  formula (\ref{eq:Wishart0}). Hence we will focus only on
the Wishart case with $M\ge N$. Note that even though the Wishart pdf
in Eq. (\ref{eq:Wishart0}) was obtained for {\em integer} $M\ge N$, the
pdf is actually a valid measure for any {\em real continuous} $M\ge N$.
In particular, for $M-N=\frac{1}{2}$ ($M-N=-\frac{1}{2}$) in the case
$\beta=2$,
the pdf in Eq. \eqref{eq:Wishart0} is realized as the distribution of the
squares
of positive eigenvalues of class $C$ ($D$) random matrices \cite{Katori2}, in the
classification of Altland and Zirnbauer \cite{Altland}. In addition, the pdf 
 in Eq. \eqref{eq:Wishart0} is also realized for $d$-dimensional squared Bessel
 processes under nonintersection constraint with $\beta=2$ and $M-N=d/2-1$
for $d>0$ \cite{Katori4}.

In Ref. \cite{Schehr}, the joint pdf of the 
positions of $N$ nonintersecting Brownian motions in the `watermelon with a
wall' geometry,  mentioned in the 
previous paragraph, 
was shown to correspond to the joint pdf of Wishart ensemble in Eq. (\ref{eq:Wishart0})    
with special values of the parameter $\beta=2$ and $M-N=1/2$.
At these special values, the joint pdf also correspond to those
of the squares of the eigenvalues of class $C$ matrices \cite{Katori2}. 
A question thus naturally arises whether it is possible to find a nonintersecting
Brownian motion model that will generate a Wishart ensemble in Eq. (\ref{eq:Wishart0}) 
with arbitrary values of the two parameters $\beta$ and $M-N\ge 0$. 
In this paper we address precisely this issue and 
show how to generate the Wishart ensemble with arbitrary positive $\beta$ and 
$M-N\ge 0$,
starting from an underlying microscopic model of nonintersecting Brownian motions.

In this paper we study the nonintersecting Brownian motions in a geometry 
different from that of 
the watermelons discussed above. Here we consider a system of $N$ 
$(1+1)$-dimensional nonintersecting
fluctuating elastic interfaces
with heights $h_i(x)$ ($i=1,2,3,\dots N)$ that run across the interval $x\in [0,L]$ in the 
longitudinal direction (see Fig. 2). Equivalently the heights $\{h_i(x)\}$ can be thought
of as the positions of $N$ nonintersecting walkers at `time' $x$.   
In contrast to the watermelon geometry, the lines 
here are not constrained to reunite at the two end points. 
Instead, each
line satisfies the periodic boundary condition in the longitudinal direction, i.e., they
are wrapped around a cylinder of perimeter $L$ (see Fig. 2). In addition, there is a hard wall (or 
substrate) 
at $h=0$ that induces an external confining potential $V(h_i)$ on the $i$-th interface for all
$1\le i\le N$. 
In a slightly more general version of the model, one can also introduce
a pairwise repulsive interaction between lines. In presence of the
external confining potential $V(h_i)$, the system of elastic lines reaches
a thermal equilibrium and our main goal is to compute the
statistical properties of the heights of these lines at thermal equilibrium.
More precisely, we compute the joint distribution of heights
of the lines at a fixed position $0\le x\le L$ and show
that, in the limit $L\rightarrow \infty$ (large system), this joint distribution, after an
appropriate change of variables, is precisely the same as
the joint distribution in the Wishart ensemble. Note that due to the translational 
symmetry
in the longitudinal direction (imposed by the periodic boundary condition), this joint pdf of the
heights is actually independent
of $x$.
\begin{figure}
\includegraphics[width=\linewidth]{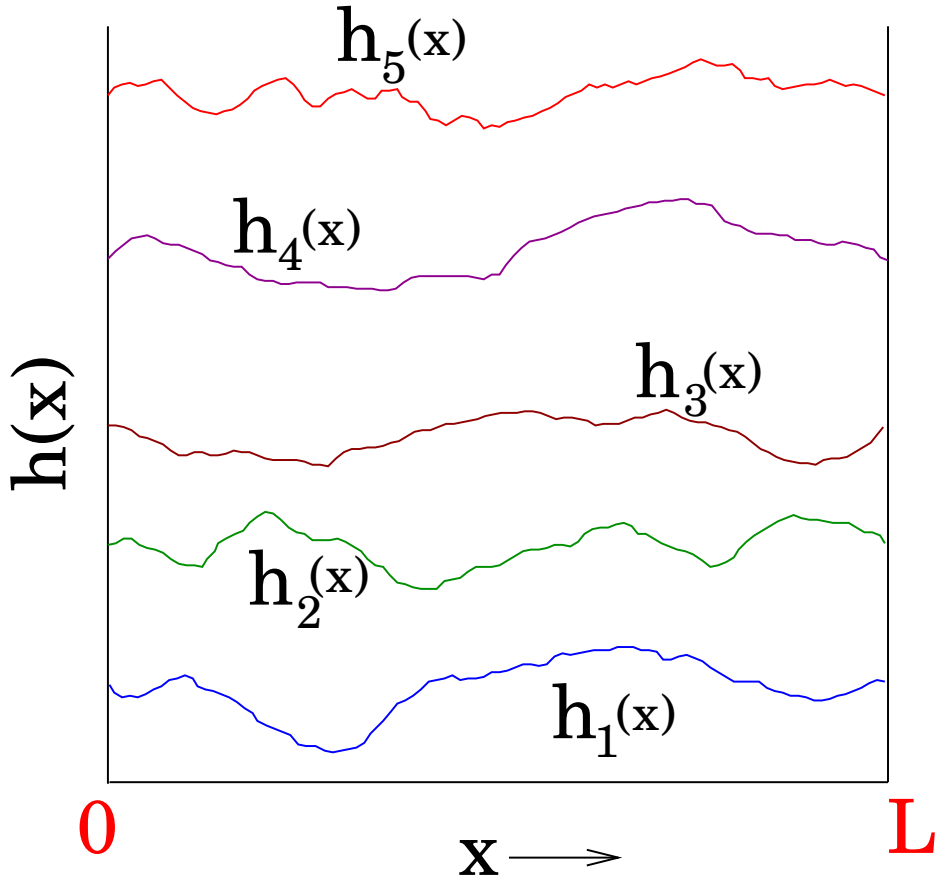}
\caption{Nonintersecting fluctuating interfaces with heights $h_i(x)$ 
for $0\le x\le L$ 
with periodic boundary conditions, $h_i(0)=h_i(L)$.}
\end{figure}

Thus our model is actually closer to the solid-on-solid (SOS) models at thermal
equilibrium in presence of a substrate~\cite{Forgacs}, except with
the difference that here we have multiple nonintersecting interfaces. 
This model is thus appropriate
to describe the interfaces between different
co-existing `wet' phases of a multiphase 
two-dimensional fluid system on a solid substrate or a film~\cite{fisher}. 
We note that nonintersecting Brownian motions in an external harmonic potential was
studied recently by Bray and Winkler~\cite{Bray}, but they were mostly interested in calculating
the probability that such walkers all survive up to some time $t$. In the Brownian
motion language, we are here interested
in a different question: given that each walker survives up to `time' $L$ and comes back to its 
starting position, what is the joint distribution of the positions of the walkers (or equivalently 
the heights of the interfaces) at any
intermediate `time' $0\le x\le L$ ? 

In this paper, we consider the external confining potential of the form      
\begin{equation}
V(h)=\frac{b^2 h^2}{2}+\frac{\alpha(\alpha-1)}{2 h^2}
\;\; \textrm{with} \;\; b>0 \;\; 
\textrm{and}\;\; \alpha >1
\label{potential}
\end{equation}
with a harmonic confining part and a repulsive
inverse square interaction. Such a choice is dictated by the
following observations. The harmonic potential is needed to confine
the interfaces as otherwise there will be a zero mode. 
The repulsive inverse square potential has an entropic origin.
For a single interface near a hard wall, 
Fisher \cite{fisher} indeed showed that the effective free energy
at temperature $T$ behaves as $k_B T/h^2$ where $h$ 
is the distance of the interface (or the walker) from the wall. 
Thus it is natural to choose the external potential of the form 
as in Eq. (\ref{potential}). In addition, as we will see later,
such a physical choice also has the advantage that it is exactly soluble.
We will see indeed that this choice of the potential generates, for the
joint density of heights at a fixed point $x$ and in the limit of a 
large system ($L\rightarrow \infty$), 
a Wishart pdf in Eq. (\ref{eq:Wishart0}) with fixed $\beta=2$, but with
a tunable $M-N= \alpha-1/2$ where $\alpha$ sets the amplitude of the repulsive
inverse square potential in Eq. (\ref{potential}).
Thus when $\alpha-1/2=k$ is a positive integer $k$, this generates a physical
realization
of Wishart matrices with integer dimensions $M=N+k$ and $N$.

Our model is also rather close to the realistic experimental
system of fluctuating step edges on vicinal surfaces of a crystal in presence
of a substrate (or hard wall).
When a crystal is cut by a plane which is oriented at a small nonzero angle to the
high-symmetry axis, one sees a sequence of terraces oriented in the high-symmetry
direction that are separated by step edges which can be modelled as `elastic'
nonintersecting lines or trajectories of nonintersecting Brownian motions~\cite{Einstein}.
In an external confining harmonic potential but in absence of a wall
at $h=0$ (such that $h\to - h$ symmetry is preserved), 
the joint distribution of the heights of the lines at equilibrium
can be mapped to the GUE ensemble~\cite{Einstein}, although most studies
in this context are concerned with the so called Terrace-Width distribution, i.e.,
the distribution of the spacings between the lines. 
In our model, due
to the presence of the wall which breaks the $h\to -h$ symmetry, new
interesting 
questions emerge. 
For example, it is natural also to ask for the 
distribution 
of the minimal (maximal) height, i.e., the height of the line closest (farthest)
from the wall. Using the mapping to the Wishart random matrix, the minimal 
and maximal height correspond respectively to the smallest and the largest
eigenvalue of the Wishart random matrix. In addition, it is also
interesting and physically relevant to investigate the statistics
of the center of mass of the $N$ nonintersecting Brownian motions.
We will see that strong correlations between the lines violate the central limit theorem
resulting in strong non-Gaussian tails in the distribution of the
center of mass. 

Let us summarize below our main results.

\vskip 0.2cm

$\bullet$ Using a path integral formalism we show that the computation of the equilibrium 
joint distribution of heights at a fixed point in space can be mapped to determining
the spectral properties of a quantum Hamiltonian. Subsequently, when the external
potential is of the form in Eq. (\ref{potential}), this quantum potential turns
out to be integrable and allows us to compute the equilibrium joint distribution 
of heights exactly. In 
particular in the limit of a large system
($L\rightarrow \infty$), we show that the joint distribution of heights, under 
a change of variables $b\, h_i^2=\lambda_i$, is exactly of the Wishart form in Eq. 
(\ref{eq:Wishart0}) with
parameters $\beta=2$ and $M-N=\alpha-1/2$ where $\alpha$ appears in the potential in Eq.
(\ref{potential}). Knowing the exact joint distribution, we then compute various
statistical properties of the heights of the interfaces as listed below.

\vskip 0.2cm

$\bullet$ We find that the average density of lines at height $h$, in the limit of
a large number $N$ of interfaces, is a quarter of ellipse as a function of $h$, with finite 
support over $[0,2 \sqrt{\frac{N}{b}}]$ where $b$ appears in Eq. (\ref{potential}). 
The typical height thus scales with $N$ for large $N$ as
$h_{\rm typ} \sim \sqrt{N}$. This differs considerably
from the case of $N$ interfaces that are 
allowed to cross, where the typical
height is of order one. The spreading of nonintersecting interfaces
is a consequence of the strong interaction between them induced 
by their fermionic repulsion.

\vskip 0.2cm

$\bullet$ We study the height distribution of the topmost (farthest from the wall) interface
in the large $N$ limit.
We show that the average height of the farthest interface (maximal height) is
$ 2 \sqrt{\frac{N}{b}}$ for large $N$ : it is
 given by the upper bound of the average density of states.
The typical fluctuations of the maximal height around its mean 
are distributed via the Tracy-Widom distribution
\cite{Johansson,Johnstone, T-W mean}. However, for finite but large $N$, the tails
of the distribution of the maximal height show significant deviations
from the Tracy-Widom behavior. We compute exactly these large deviation tails.

\vskip 0.2cm

$\bullet$ We also study the statistics of the height
of the lowest (closest to the wall) interface (minimal height) and argue
that, for large $N$, it scales as $N^{-1/2}$. This should be compared
to the case of non-interacting Brownian motions where the typical
distance of the closest (to the wall) walker is of $\sim O(1)$ from
the wall. 
This is again an effect of the strong interaction
between the interfaces: their mutual repulsion pushes the lowest interface closer
to the substrate.
We further show that the full distribution
of the minimal height can be exactly computed
for a special value of the parameter $\alpha=3/2$ in Eq. (\ref{potential}).

\vskip 0.2cm 

$\bullet$ Finally we study the distribution of
the center of mass of the heights $G_N=\frac{h_1+...+h_N}{N}$ for large $N$.
Thanks to the analogy between the Wishart eigenvalues and a Coulomb gas
of charges, the mean and variance
of the center of mass can be computed,
 as well as the shape of the  probability distribution: we show that
the pdf of $G_N$, 
 $P(G_N=\nu)$, has a non-analytic behavior 
(essential singularity) at $\nu=\langle G_N \rangle$
(which is shown to be a direct consequence of a phase transition of `inifinite' order in the associated 
Coulomb
gas problem).
In addition, we find exact asymptotic results, to leading order for large $N$,
for the mean $\langle G_N \rangle=\frac{8}{3 \pi} \sqrt{\frac{N}{b}}$ 
and the variance $\langle G_N^2 \rangle -\langle G_N \rangle^2 =\frac{2}{\pi^2 N b}$.

The rest of the paper is organized as follows.
In section \ref{sec:model}, we present our model, compute
(via path integral method)
the joint probability distribution of the
heights of the interfaces and compare it to the probability distribution
of the eigenvalues of a Wishart matrix.
In section \ref{sec:StatProp}, we analyse some statistical properties
of the model. We first present the results for the average density of lines
(subsection \ref{subsec:dens}).
We then compute the behavior of the maximal height 
(\ref{subsec:max}) and the minimal height (\ref{subsec:min}).
Finally we  study the distribution of the
center of mass of the heights in (\ref{subsec:cdm}).
Section \ref{sec:ccl} concludes the paper with a summary and outlook.

\section{The model}
\label{sec:model}

Our model consists of $N$ nonintersecting $(1+1)$-dimensional
 interfaces over a substrate of size $L$ (that induces an external potential).
For simplicity, we first present the model for a single interface in subsection \ref{1 int}
and show how, using a path-integral formalism, one can map the problem of calculating the equilibrium 
height distribution of the interface to computing the spectral properties 
of a quantum Hamiltonian. In particular, calculating the height distribution
in the limit $L\rightarrow\infty$ (large system)
 corresponds to calculating the ground state wavefunction of
this quantum Hamiltonian. Then we present the interacting model for general $N$
interfaces in subsection \ref{N int}
 and show how to generalize the path integral formalism
to a {\em many-body} problem and subsequently compute the joint distribution 
of heights at equilibrium.
In particular,
for a large system,
 the joint distribution is shown to have the Wishart form in Eq. (\ref{eq:Wishart0}) 
with
parameters $\beta=2$ and $M-N=\alpha-1/2$ where $\alpha$ appears in the potential in Eq. 
(\ref{potential}).  

\subsection{One interface}
\label{1 int}

Let us first consider the case of one single interface ($N=1$).
The interface is described by its height $h(x)$ for $x$ from $0$ to $L$.
When we think of the interface as a walker (or Brownian motion),
the height $h$ plays the role of the position of the walker,
while the coordinate $x$ along the substrate corresponds to  time.
The substrate can  then be seen as a wall at height zero:
one has $h(x)>0$ for every $x$.
In the stationary state at thermal equilibrium,
the energy of a configuration  $\{h(x)\}$
of the interface can be expressed as 
\begin{equation}
\label{eq: energy}
E \big[\{h(x)\}\big]=E_{elast}\big[\{h(x)\}\big]+U\big[\{h(x)\}\big]
\end{equation}
 with
$E_{elast}\big[\{h(x)\}\big]=\frac{1}{2} \int_0^L dx\, 
 \Big( \frac{dh}{dx} \Big)^2 $
being the elastic energy 
(or the kinetic energy of the walker) and $U\big[\{h(x)\}\big]=\int_0^L dx \,
V(h(x))$ describes the potential energy due to the interaction potential
$V(h)$ with the substrate.  
The statistical weight of a configuration $[\{ h(x)\},\,
0\leq x \leq L ]$ of the interface is thus simply (setting $k_B T =1$ where $k_B$
is the Boltzmann constant and $T$ the temperature) given by the Boltzmann weight
\begin{equation}
P\left[ \{ h(x) \} \right]  \propto  \exp \big\{-E \big[\{h(x)\}\big]
\big\}\propto  \exp\left\{-    \frac{1}{2} 
\int_0^L \Big( \frac{dh}{dx} \Big)^2 dx
-   \int_0^L  V\left(h(x)\right) dx\right\} 
\label{boltzmann}
\end{equation}
We assume periodic boundary conditions: $h(0)=h(L)=h >0$.

In absence of an external potential ($V \equiv 0$), the interface is depinned, with a
 roughness exponent $\chi=1/2$, which means
$\langle h^2 \rangle -\langle h \rangle^2 \propto L^{2 \chi} \propto L$.
In this case the interface is just the trajectory of a free one dimensional Brownian motion:
the displacement $h$ of the walker grows as the square root of the `time'
($L$). But when the substrate induces an attractive potential,
the interface remains pinned to the wall (substrate).
The interface then becomes smooth ($\chi=0$) in this case.

Given the overall statistical weight of the {\em full configuration} of an interface
in Eq. (\ref{boltzmann}) over $x\in [0,L]$,
our task next is to compute the `marginal' height distribution $P(h)$ of the interface at a
fixed point $x$ in space, by integrating out the heights at other points.
Note that due to the translational symmetry imposed by the periodic boundary condition,
this marginal height distribution $P(h)$ is independent of the point $x$, which
we can conveniently choose to be $x=0$ for example.
This integration of all other heights except at $0$ (or $L$) can be very conveniently carried
out by the following path integral that allows us to write the marginal pdf $P(h)$
as
\begin{equation}
P(h) \propto \int_{h(0)=h}^{h(L)=h}  \mathcal{D}h(x)
\:
e^{ -E \big[\{h(x)\}\big] }\,\,
\mathbb{1}_{h(x)>0 }
\label{Ph0}
\end{equation}
where the symbol $\mathbb{1}_{h(x)>0}$ is an indicator function
that enforces the condition that the height at all points $x\in [0,L]$ is positive
and the energy $E\big[\{h(x)\}\big]$ is given in Eq. (\ref{boltzmann}).

The path integral can be reinterpreted as a quantum propagator:
\begin{equation}
P(h) \propto \langle h | e^{-\hat{H} L} |h\rangle
\label{Ph1} 
\end{equation}
with the Hamiltonian
\begin{equation}
\hat{H}=- \frac{1}{2} 
 \frac{d^2 }{dh^2} +  V(h)
\label{Hamil0} 
\end{equation}
and with the constraint $ h>0$.
The problem is now the one of a quantum particle
in one dimension
with  position $h(x)$ at time $x$, described by
the  Hamiltonian $\hat{H}$
(in imaginary time).

We assume now that the energy spectrum of $\hat{H}$ is discrete 
(this will be the case in presence of the confining potential).
The propagator can be decomposed in the eigenbasis of $\hat{H}$:
\begin{equation}
P(h)=\frac{\sum_E e^{-E L} \:
 \left| \psi_E(h) \right|^2}{ \sum_E e^{-E L}}
\label{marginal1}
\end{equation}
where $\psi_E$ is the eigenfunction of energy $E$.
Thus calculating the marginal height distribution is equivalent, thanks
to the relation in Eq. (\ref{marginal1}), to calculating the full
spectral properties (i.e., all eigenvalues and eigenfunctions) of
the quantum Hamiltonian $\hat{H}$.
In Eq. \eqref{Ph1} and \eqref{marginal1}, the size  $L$ of the substrate
(in the classical system of interfaces)
plays the role of the inverse temperature in the associated quantum problem
(but it has nothing to do with the temperature of the interfaces).
Hence, in a large system $L\to \infty$, 
only the ground state ($\psi_{E_0} \equiv \psi_0$ with energy $E_0$)
contributes to the sum in Eq. (\ref{marginal1}). 
Henceforth, we will always work in this limit 
where the marginal pdf is given by the exact formula
\begin{equation} 
P(h)=|\psi_0(h)|^2
\label{marginal2}
\end{equation}

Let us make a quick remark here. While the results in Eqs. (\ref{marginal1}) and (\ref{marginal2})
may apriori look evident, they are however a bit more subtle. For example, the r.h.s.
of Eq. (\ref{marginal2}) is, in a quantum mechanical sense, the probability density
of finding a particle at $h$ in the ground state. But it is not obvious (and needs
to be proved as done above) that it also represents the height distribution
of a classical model. 

Thus our task is now to determine the exact ground state of the 
quantum Hamiltonian $\hat H$. Analytically this is only possible 
for integrable $\hat H$. With the choice of potential $V(h)$
as in Eq. (\ref{potential}), the quantum Hamiltonian $\hat H$
is fortunately integrable.
The eigenfunction $\psi_n(h)$ satisfies 
the Schr\"odinger equation:
\begin{equation}
\hat{H} \psi_n= -\frac{1}{2} 
 \frac{d^2 \psi_n}{dh^2} +  V(h) \psi_n = E_n \psi_n
\label{Shrod1} 
\end{equation}
with the boundary conditions, $\psi_n(h=0)=0$ (due to the hard wall at $h=0$)
and $\psi_n(h\to \infty)=0$. The solution is of the form
\begin{equation}
\label{SinglePartic}
\psi_n(h) =c_n \: e^{-\frac{b}{2} h^2 } 
\; h^{\alpha} \;
\mathcal{L}_n^{(\alpha-\frac{1}{2})}(b h^2); \quad\quad
\textrm{with}\: \;\; 
E_n=b\,(2 n +\alpha +\frac{1}{2})
\end{equation} 
with $n$ a non-negative integer (discrete spectrum),
 $c_n$ a normalization constant and $\mathcal{L}_n^{(\alpha-\frac{1}{2})}$
a generalized Laguerre polynomial of degree $n$
\begin{equation}
\mathcal{L}_n^{\gamma}(x)=\sum_{i=0}^n 
\left( \begin{array}{c}
n+\gamma\\ n-i \end{array} \right)\, \frac{(-x)^i}{i!}
\label{laguerre1}
\end{equation}
Note that for $\gamma=0$, $\mathcal{L}_n^{0}(x)=\mathcal{L}_n(x)$ reduces
to the ordinary Laguerre polynomial
\begin{equation}
\mathcal{L}_n(x)= \sum_{i=0}^n
\left( \begin{array}{c}
n\\ i \end{array} \right)\, \frac{(-x)^i}{i!}.
\label{laguerre0}
\end{equation}
We also note that the generalized Laguerre polynomial in Eq. (\ref{laguerre1}) can alternately be 
expressed as a
hypergeometric function
\begin{equation}
\mathcal{L}_n^{\gamma}(x)= \frac{(\gamma+1)_n}{n!} \, _1F_1(-n; \gamma+1; x).
\label{hyper1}
\end{equation}
where $(a)_n=(a) (a+1) ... (a+n-1)$ is the Pochhammer symbol and 
\begin{equation}
_pF_q(a_1,...,a_p ; b_1 ,..., b_q ; z)=
\sum_{n=0}^{\infty} \frac{(a_1)_n ... (a_p)_n}{(b_1)_n  ... (b_q)_n}
\frac{z^n}{n!}.
\label{hyper2}
\end{equation}

Finally, as $\mathcal{L}_0^{\gamma}(x)= 1$, 
the pdf of the height of the interface is
(in the limit of a large substrate $L \rightarrow \infty$)
\begin{equation}
P(h)=|\psi_0(h)|^2=|c_0|^2 \,  \: e^{-b \, h^2 } 
\; h^{2 \alpha}
\label{heightd1} 
\end{equation}
with $|c_0|^2 =  \frac{2\, b^{\alpha+1/2}}{\Gamma(\alpha +1/2)} $.

The mean $m\equiv \langle h \rangle$ and variance 
$\sigma_1^2 \equiv {\rm Var}(h)=\langle h^2 \rangle-\langle h \rangle^2$
of the interface height are thus easy to compute:
\begin{eqnarray}
&&
\hspace{-0.5cm} m=\langle h \rangle=\int_0^{\infty} dh \, h\, P(h)=
\frac{\Gamma(\alpha +1)}{\sqrt{b}\;\, \Gamma(\alpha+1/2)}\label{mean1} \\
&& \hspace{-0.5cm} 
\sigma_1^2={\rm Var}(h)=\frac{1 +2 \alpha}{2 \, b}-
\frac{1}{b}\, 
\left( \frac{\Gamma(\alpha +1)}{ \Gamma(\alpha+1/2)} \right)^2 \label{var1}
\end{eqnarray}

\subsection{$N$ interfaces}
\label{N int}

Let us consider now  $N$ nonintersecting $(1+1)$-dimensional
 interfaces over a substrate of size $L$.
The $i^{th}$
 interface is described by its height $h_i(x)$ for $x$ from $0$ to $L$.
Since the interfaces are nonintersecting, we can assume
that they are ordered: $0<h_1(x)<h_2(x)<...<h_N(x)$ for every $x$.
The only interaction between the interfaces is
their fermionic repulsion (they do not cross).
However, we will see that this constraint 
drastically changes the statistics of the interfaces.
In the stationary state (at thermal equilibrium),
 the energy  $E \big[\{h(x)\}\big]$ of a configuration  $\{h(x)\}$
of one of the interfaces is given by \eqref{eq: energy} (same form as
we assumed for one single interface),
with an elastic energy and a potential 
$V(h)$ again given by \eqref{potential}.
Therefore the
 statistical weight of a configuration $\{ h_i(x)\,  ; 1 \leq i \leq N, \:
0\leq x \leq L \}$ of the whole system is  simply 
(setting $k_B T=1$ for simplicity):
\begin{equation}
P\left[ \{ h_i(x)  \}_{i, x} \right]
  \propto \exp\left[- \sum_i \frac{1}{2} 
\int_0^L \Big( \frac{dh_i}{dx} \Big)^2 dx   -\sum_i \int_0^L  V\left(h_i(x)\right) dx  \right] 
\label{energy1}
\end{equation}
We  assume again  periodic boundary conditions:
for every $i$, $h_i(0)=h_i(L)=h_i$ with 
$0<h_1<h_2<...<h_N$. The configuration space can thus be seen as a cylinder
of radius $L/2 \pi$.

The joint probability distribution of the heights of the interfaces at a
given position (position $x$ that can be taken to be $0$ by cylindrical 
symmetry, as we already noticed) can again be expressed as a path integral:
\begin{equation}
P(h_1,h_2,...,h_N) \propto
\prod_i \int_{h_i(0)=h_i}^{h_i(L)=h_i}  \mathcal{D}h_i(x)
\:
e^{ -\sum_i E \big[\{h_i(x)\}\big] }\:
\mathbb{1}_{ h_N(x)>...>h_1(x)>0 }
\label{Njpdf1}
\end{equation}

The path integral can then be reinterpreted as a quantum propagator
for $N$ particles:
\begin{equation}
P(h_1,h_2,...,h_N) \propto
\langle h_1,h_2,...,h_N | e^{-\hat{H} L} |h_1,h_2,...,h_N \rangle
\label{Nprop1} 
\end{equation}
where the many-body Hamiltonian is given by
\begin{equation}
\textrm{with } \:
\; \hat{H}=\sum_i \hat{H}_i=-\sum_i \frac{1}{2} 
 \frac{d^2 }{dh_i^2} + \sum_i V(h_i)
\label{NHamil1} 
\end{equation}
with the constraint $  h_N>...>h_1>0$.
The problem is now the one of $N$ independent fermionic particles
in one dimension
with positions $h_i(x)$ at time $x$, described by
the single particle Hamiltonian $\hat{H}_i$
(in imaginary time).

Exactly as for one single interface,
the propagator can be decomposed in the eigenbasis
of $\hat{H}$ (Hamiltonian for $N$ particles) and
the joint distribution of heights is given by
\begin{equation}
P(h_1,h_2,\ldots,h_N)= \frac{\sum_E e^{-E L} \:
 \left| \psi_E(h_1,h_2,\ldots,h_N) \right|^2}{ \sum_E e^{-E L}}
\label{mbmarginal1}
\end{equation}
where $\psi_E(h_1,h_2,\ldots,h_N)$ is the many-body wavefunction at energy $E$.
Analogus to the single interface case, 
when the size of the system $L$ tends to infinity,
only the  ground state $ \Psi_0$ ($N$-body wavefunction)
contributes to the sum. In this limit, the joint probability is simply:
\begin{equation} 
P(h_1,h_2,...,h_N)=|\Psi_0(h_1,...,h_N)|^2
\label{mbmarginal2}
\end{equation}
As in the single particle case, we emphasise that the relations in Eqs. (\ref{mbmarginal1})
and (\ref{mbmarginal2}) may apriori look evident, but they need to be proved
as the l.h.s and r.h.s. of these equations refer to the
probability density in a classical and a quantum problem respectively.  
We note that in the context of step edges on vicinal surfaces (in the absence
of a wall), the relation (\ref{mbmarginal2}) was implicitly assumed in Ref. \cite{Einstein}, but
not proved.

In the case of one interface, we computed the
single  particle wavefunction $\psi_n$ in \eqref{SinglePartic}.
As the particles are independent fermions, the ($N$-body) ground state
wavefunction $\Psi_0$ is a  $N \times N$ Slater determinant.
It is constructed from the $N$  single particle wavefunctions
of lowest energy, the $\psi_i$ for $i$ from $0$ to $N-1:$
\begin{equation}
\Psi_0(h_1,...,h_N) \propto \det \left( \psi_{i-1} (h_j) \right) 
\propto
 e^{-\frac{b}{2}  \sum_k h_k^2 } 
\; \prod_k h_k^{\alpha} \;\:
 \det\left(
\mathcal{L}_{i-1}^{(\alpha-\frac{1}{2})}(b h_j^2)
 \right)
\label{slater1}
\end{equation}

Note that $\mathcal{L}_{i-1}^{(\alpha-\frac{1}{2})}(b h^2)$
 is a polynomial of $h^2$ of degree $i-1$.
Any determinant involving polynomials can be reduced, via the linear combination
of rows, to a Vandermonde determinant which can then be simply evaluated.
We then get
\begin{equation}
P(h_1,...,h_N)=|\Psi_0(h_1,...,h_N)|^2\, \propto
 e^{- b  \sum_k h_k^2 } 
\; \prod_k h_k^{2 \alpha} \;\:
\prod_{i<j} (h_i^2-h_j^2)^2 
\label{jpdfh1}
\end{equation}
where $h_i$'s are positive. Note that
due to the symmetry of the above expression,
the ordering constraint $h_1<...<h_N$ can be removed   
by simply dividing the normalization constant
by $N!$.

For  interfaces that are allowed to cross,
the joint probability distribution of the heights
has a similar form, but
without the Vandermonde determinant:
$P(h_1,...,h_N)=P(h_1) ... P(h_N)  \propto
 e^{- b  \sum_k h_k^2 } 
\; \prod_k h_k^{2 \alpha}$. 
The Vandermonde determinant $\prod_{i<j} (h_i^2-h_j^2)^2$
comes from the fermionic repulsion between the interfaces.
In particular, one has (as expected) for nonintersecting interfaces
$P(h_1,...,h_N)=0$ if $h_i=h_j$ for $i \neq j$. The consequence
of this repulsion on the typical magnitude of the heights of
interfaces will be explored in the next section. 

\subsection{Relation to the Eigenvalues of a Wishart Matrix}
\label{evWishart}

We recall from the introduction that an $(N\times N)$ Wishart matrix is a product
covariance matrix of the form
$W=X^{\dagger}X$ where $X$ is a Gaussian $(M\times N)$ rectangular matrix drawn from
the distribution, $P(X)\propto \exp\left[-\frac{\beta}{2} {\rm
Tr}(X^{\dagger} X)\right]$. For $M\ge N$, all eigenvalues of $W$ are non-negative
and are distributed via the joint pdf in Eq. (\ref{eq:Wishart0}).
In the ``Anti-Wishart'' case, that is when $M<N$,
$W$ has $M$ positive eigenvalues ( and $N-M$ eigenvalues
that are exactly zero ) and their 
joint probability distribution is simply obtained by
exchanging $M$ and $N$
in the  formula (\ref{eq:Wishart0}).

The joint probability distribution  of the heights of the interfaces
in our model in Eq. (\ref{jpdfh1}) can then be related
to the Wishart pdf in Eq. (\ref{eq:Wishart0}) with $\beta=2$ after a
change of variables $b\, h_i^2=\lambda_i$
\begin{eqnarray} 
P(h_1,\ldots,h_N) dh_1 ... dh_N  
&& \propto  e^{- b  \sum_k h_k^2 } 
\; \prod_k h_k^{2 \alpha} \;\:
\prod_{i<j} (h_i^2-h_j^2)^2
dh_1\ldots dh_N \nonumber \\
&& = A_N\,  e^{-  \sum_k \lambda_k } 
\; \prod_k \lambda_k^{ \alpha- \frac{1}{2}} \;\:
\prod_{i<j} (\lambda_i-\lambda_j)^2
d\lambda_1 \ldots d\lambda_N  
\label{jpdfh2}
\end{eqnarray}
where $A_N$ is a normalization constant.
Recall that the parameter $\alpha>1$.
By choosing $\alpha= (M-N)+\frac{1}{2}>1$,
one recovers the Wishart pdf in Eq. (\ref{eq:Wishart0})
for $\beta=2$  (complex matrices $X$) and arbitrary $M-N>1/2$. 
Thus by tuning the amplitude $\alpha$ of the repulsive
part of the potential in Eq. (\ref{potential}) one
can generate the Wishart ensemble with a tunable $M$ (with $M-N>1/2$).
The normalization constant $A_N$ can be computed
using Selberg's integrals \cite{Mehta} and one gets
\begin{equation}
  A_N^{-1}= 
 \prod_{k=1}^N \left( k! \,\: \Gamma(\alpha -\frac{1}{2}+k)\right)
\label{normal1}
\end{equation}
\\

\noindent {\bf Wishart pdf with arbitrary $\beta\ge 0$ and $M> N$:} We note that
our model above generates a Wishart pdf with arbitrary $M-N>1/2$ but with
fixed $\beta=2$. It is possible to generate the Wishart pdf with arbitrary $\beta\ge 0$
also by introducing an additional pairwise repulsive potential between the interfaces.
For example, we may add to the energy functional in Eq. (\ref{energy1}) an
additional pairwise interaction term of the form,
\begin{equation}
-\sum_{1\le j<k\le N}\,\int_0^L V_{\rm pair}\left(h_j(x),\,h_k(x)\right)\, dx
\label{pairpot1}
\end{equation}
where the pair potential $V_{\rm pair}(h_j,h_k)$ has a specific form
\begin{equation}
V_{\rm pair}(h_j,h_k)= 
\frac{\beta}{2}\,\left(\frac{\beta}{2}-1\right)\,\left[\frac{1}{(h_j-h_k)^2}+\frac{1}{(h_j+h_k)^2}
\right]
\label{pairpot2}
\end{equation}
with $\beta\ge 0$.
In this case, once again using the path integral formalism developed above, we can
map the computation of the joint distribution of heights to calculating
the spectral properties of a quantum Hamiltonian via Eqs. (\ref{mbmarginal1})
and (\ref{mbmarginal2}). The corresponding quantum Hamiltonian turns out 
to be exactly the Calogero-Moser model~\cite{Moser} which is integrable~\cite{OP,YKY}. 
In particular, using the exact ground state wavefunction of this Hamiltonian 
we get our corresponding joint distribution of interface heights 
in the following form
\begin{equation}
P(h_1,...,h_N)=|\Psi_0(h_1,...,h_N)|^2 \propto
 e^{- b  \sum_k h_k^2 }
\; \prod_k h_k^{2 \alpha} \;\:
\prod_{i<j} (h_i^2-h_j^2)^{\beta}
\label{varbeta1}
\end{equation}
which, after the usual change of variables $b\, h_i^2=\lambda_i$, corresponds
to the general Wishart pdf in Eq. (\ref{eq:Wishart0}) with
arbitrary $\beta\ge 0$ and a tunable $M-N=(2\alpha+1-\beta)/\beta$.
We note that the procedure used above to obtain a variable $\beta$
random matrix ensemble was used before in the context of step edges on vicinal surfaces
{\em without a hard wall} where a corresponding Gaussian matrix ensemble
with tunable $\beta$ was obtained~\cite{Einstein}. 

Note that for interfaces in presence of a wall, while the first term in the pair potential in Eq. 
(\ref{pairpot2})
is quite natural and can arise out of entropic origin as well as dipolar interaction between step 
edges~\cite{Einstein}, the second term however does not have any physical origin.
Unfortunately if one gets rid of this term, the integrability of the quantum 
Hamiltonian also gets lost.
In any case, in the following we would focus only on the physical $\beta=2$ case.

\section{Statistical properties of the model}
\label{sec:StatProp}

Once the joint distribution of heights at equilibrium is known, one can,
at least in principle, compute the 
statistics of various relevant quantitites such as the average density of lines at height $h$,
the distribution of the maximal and the minimal height, the distribution
of the center of mass of the interfaces etc. In this section
we show how to carry out this procedure and derive some explicit results upon
borrowing the techniques developed in the context of random matrix theory.
Many of the results in Secs. III A-C actually follow from a simple change of
variables in the already known results of random matrix theory. For the sake
of completeness, we remind the readers some of these results from random
matrix theory
 and draw the
consequences
 for our interface model. However, Sec. III D (where we derive the
 distribution of the center of mass of the interfaces) presents completely new results.

For simplicity, we will focus here only on large system ($L\rightarrow
\infty$)
properties.
In particular, our focus would be to understand
the effect of fermionic repulsion between the interfaces 
(nonintersecting constraint) and also the effect of the external confining
potential on the statistics of the above mentioned physically relevant quantities.

We have shown in the previous section that the joint pdf 
of interface heights ($h_i$), after the change of variables
$b\, h_i^2=\lambda_i$,
is the same as the joint pdf of Wishart eigenvalues
($\lambda_i$) in Eq. (\ref{eq:Wishart0}), which can be re-written
as a Boltzmann weight
\begin{equation}
P_N(\lambda_1,...,\lambda_N)\propto \exp \left[ - \beta E_{\rm eff}(\lambda_1, ..., \lambda_N) 
\right]
\label{boltzmann2}
\end{equation}
with the effective energy 
\begin{equation}
E_{\rm eff}= \frac{1}{2}\,  \sum_k \lambda_k-a 
 \sum_k \ln \lambda_k - \sum_{i<j} \ln |\lambda_i -\lambda_j|
\label{energy2}
\end{equation}
where $a=\left(\frac{1+M-N}{2}-\frac{1}{\beta} \right)$. 
In this form, the $\lambda_i$'s can be interpreted as the positions 
of charges repelling each other via the $2$-d Coulomb interaction (logarithmic),
but are confined on the $1$-d positive axis and in presence of an external
linear+logarithmic potential.  
The Dyson index $\beta$ 
plays the role of inverse temperature.
Our model of interfaces corresponds to $\beta=2$ (see Eq. \eqref{jpdfh2}).

We already noticed that the nonintersection constraint for the interfaces
is equivalent to the presence of the Vandermonde determinant, and thus
the logarithmic Coulomb repulsion, in the joint probability distribution.
For independent interfaces (allowed to cross), there is no Vandermonde term
and hence the logarithmic 
repulsion term in Eq. (\ref{energy2}) is absent.
In that case,
balancing the first two
terms of the energy gives a typical height
 of order one: $h_{\rm typ} \sim O(1)$. But for the nonintersecting case,
when the number $N$ of interfaces becomes large,
the logarithmic repulsion is stronger than the logarithmic part of
the external potential (provided $a$ is not proportional to $N$).
Therefore, balancing the first and the third term in  
the effective energy gives, for large $N$, $N \lambda_{\rm typ} \sim N^2$, thus
$\lambda_{\rm typ} \sim N$ or equivalently $h_{\rm typ} \sim \sqrt{N}$.
The effect of repulsion is strong: the interfaces spread out considerably.

Below we first compute the average density of states, followed by
the computation of the 
distribution of the topmost interface (maximal height) and the lowest interface 
(minimal height) that is the closest to the substrate. Finally
we analyze the distribution of the center of mass of the heights.

\subsection{Average density of states}
\label{subsec:dens}

We would first like to know what fraction of $N$ interfaces lie,
on an average, 
within a small interval of heights $[h,h+dh]$. This is given 
by the average density of states (normalized to unity)
\begin{equation}
\rho_N(h)=\frac{1}{N} \sum_{i=1}^N \langle \delta(h-h_i) \rangle
\end{equation}
As we explained above, we expect the typical height scale to
be of order $h_{\rm typ} \sim \sqrt{N}$ for a large number $N$ of interfaces.
Furthermore, the density of states is normalized to unity:
$\int_0^{\infty} dh \: \rho_N(h)=1$. Therefore the density is expected 
 to have the following
scaling form for large $N$:
\begin{equation}
\label{eq:density}
\rho_N(h) \approx   \frac{1}{\sqrt{N} } \;\: g\left( \frac{h}{\sqrt{N}}\right)
\end{equation}

Our goal is to compute this scaling function $g(x)$. This can actually be simply read off
from the known results on Wishart matrices which we now recall.
Consider the Wishart matrix with $M\geq N$ with eigenvalues distributed via
the joint pdf in Eq. (\ref{eq:Wishart0}).
In the asymptotic limit
$N\to \infty$, $M\to \infty$ keeping the ratio $c=N/M$ fixed (with $c\le 1$),
the average density of states of the 
eigenvalues
is known~\cite{marcenko} to be of the form:
\begin{equation}
\label{marcScaling}
\rho_N^W(\lambda)=\frac{1}{N} \sum_{i=1}^N \langle \delta(\lambda-\lambda_i)
 \rangle \approx \frac{1}{N} f\left( \frac{\lambda}{N}\right)
\;\; \textrm{for large $N$}
\end{equation}
where the  Mar\u cenko-Pastur scaling function $f(x)$ 
depends on $c$ (but is independent of 
$\beta$) 
\begin{equation}
f(x)=\frac{1}{2 \pi x} \sqrt{(x_+-x)(x-x_-)} 
\label{marcenko1} 
\end{equation}
which has a non-zero support over the interval $x\in [x_-,x_+]$ where
$x_{\pm}=\left(\frac{1}{\sqrt{c}} \pm 1 \right)^2$. Note that in
the limit $c\to 1$, which happens when $M-N\sim O(1)$ for large $N$,
$x_{-}\to 0$ and $x_{+}\to 4$. 

\begin{figure}[htp]
\centering
\begin{pspicture}(-1,0)(6,6)
\hspace{-1.5cm}
\includegraphics[width=8.4cm]{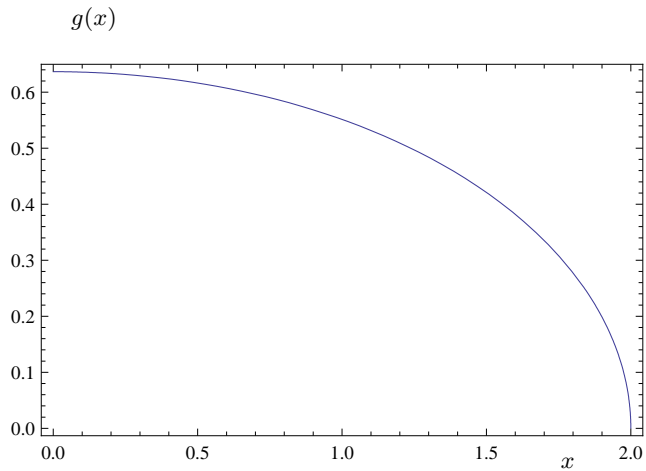}
\rput{0}(-7.3,5.9){$g(x)$}
\rput{0}(-1,0){$x$}
\end{pspicture}
\caption{Rescaled average density of states $g(x)$
for the heights of interfaces plotted for $b=1$:
$\rho_N(h) \approx   \frac{1}{\sqrt{N} } \;\: 
g\left( \frac{h}{\sqrt{N}}\right)$. It is a quarter
of ellipse.}\label{fig:dens}
\end{figure}

Our interface model, after the customary change of variable $b\,h_i^2=\lambda_i$,
corresponds to the Wishart ensemble in Eq. (\ref{eq:Wishart0}) with
$\beta=2$ and $M-N = \alpha-1/2$. Hence, as long as $\alpha\sim O(1)$
for large $N$, $c=N/M\to 1$ in our model.
Using $bh^2=\lambda$ and $c=1$ in Eq. \eqref{marcScaling}
and (\ref{marcenko1}),
the average density of 
states in
the interface model then indeed has the scaling form in Eq. (\ref{eq:density}) for large $N$ 
with the scaling function 
\begin{equation}
 g(x)=\frac{b}{\pi} \sqrt{\frac{4}{b}-x^2}
\end{equation}
where $b$ is the frequency of the harmonic part of the potential 
(see \eqref{potential}).
The average density is a quarter of ellipse, as shown in figure \ref{fig:dens}.
 It has a finite support
$\left[0\, ,\,  2\, \sqrt{\frac{N}{b}} \right]$.

Thus the interface heights spread out for large $N$ 
as a result of the nonintersection constraint.
Let us compare this result to the case of independent interfaces
that are allowed to cross each other.
In that case, the average density of states
is simply $\rho_N(h)=P(h) \propto e^{-b \, h^2 } \; h^{2 \alpha} $.
It is independent of $N$ (evidently!) and has a non-zero support over the whole
positive $h$ axis. It
vanishes when $h$ tends to zero, and rapidly decreases
to zero when $h$ becomes large. Thus most of the interfaces
lie on an average close to the wall at a distance of $O(1)$. 
In contrast, the heights of nonintersecting interfaces, on an average for large $N$,
have a compact support over a wide region. 
The density vanishes at the upper edge as a square root 
singularity and the upper edge 
itself grows as 
$\sqrt{N}$, thus spreading the interfaces further and further away from the wall
as $N$ increases.

The average of the upper height (maximum) is  expected 
to be given by the upper bound of the density support:
$\langle h_{\rm max}\rangle  \approx \:\, 2\, 
\sqrt{\frac{N}{b}}$.
The lower bound of the support of the density
is zero for large $N$ in first approximation. We will show more precisely
that the average height of the lower interface (minimum) is
proportional to
$ \frac{1}{ \sqrt{b  N}}$.

Finally, the average of all interface heights $\langle h\rangle =\langle \frac{(h_1+h_2+\ldots 
+h_N)}{N} \rangle$ can be computed for large $N$:
\begin{equation}
\label{meanh}
\langle h \rangle= \int_0^{\infty} h \rho_N(h) dh
\approx \frac{8}{3 \pi }\,  \sqrt{\frac{N}{b}}
\end{equation}
This differs drastically from the case of independent interfaces, 
where the average of all heights
is the same as that for one single interface:
$\langle h \rangle=\frac{\Gamma(\alpha +1)}{\sqrt{b}\;\,
 \Gamma(\alpha+1/2)}$ (see section \ref{1 int})
is independent of $N$, but depends on
$\alpha$ (parameter associated to
the  part of the potential proportional to $\frac{1}{h^2}$).
In contrast, for nonintersecting interfaces,
the  part of the potential proportional to $\frac{1}{h^2}$
becomes negligible compared to
the repulsion between interfaces, thus $\langle h \rangle$
does not depend on $\alpha$, but grows with $N$ (see \eqref{meanh}).
\\

\subsection{Maximal height of the interfaces}
\label{subsec:max}

In this subsection we compute the distribution of the
height of the topmost interface (maximal height),
the one furthest from the substrate.
The average of the maximal height
is given by the upper bound
of the density support
(see section \ref{subsec:dens}):
\begin{equation}
\langle h_{\rm max}\rangle  \approx \:\, 2\, 
\sqrt{\frac{N}{b}} \;\; \textrm{for large $N$.}
\end{equation}
But we would like to know the full distribution
of the height of the topmost interface, not just its average.
For that purpose, we can again take advantage of the mapping
between our interface model and the Wishart random matrix.
Under this mapping, the height of the topmost interface
$h_{\rm max}$ is related, via the change of variable $b\,h_{\rm max}^2=\lambda_{\rm max}$,
to the largest eigenvalue $\lambda_{\rm max}$ of the Wishart matrix.
The distribution of $\lambda_{\rm max}$ has been studied
in great detail and we can then directly use these results
for our purpose.

Let us recall briefly the known properties of the largest eigenvalue $\lambda_{\rm max}$
of Wishart matrices whose eigenvalues are distributed via the pdf in
Eq. (\ref{eq:Wishart0}). For our purpose we will only focus on $\beta=2$
and $M\ge N$ with $M-N\sim O(1)$. In this case the parameter
$c=N/M$ tends to the limiting value $c=1$ for large $N$, indicating
that the upper edge of the Mar\u cenko-Pastur sea, describing the average density of states
in Eq. (\ref{marcenko1}), 
approaches $x_{+}\to 4$
and lower edge $x_{-}\to 0$. Thus,
the average of the maximal eigenvalue of a Wishart matrix
is $\langle \lambda_{\rm max} \rangle \approx 4 N$.
Furthermore, Johansson \cite{Johansson} and Johnstone \cite{Johnstone}
independently showed that the typical fluctuations of $\lambda_{\rm max}$
around its mean $4N$ are of order $N^{1/3}$, i.e.,
\begin{equation}
\lambda_{\rm max} \to 4N + 2^{4/3}\, N^{1/3}\, \chi_2
\label{typ1}
\end{equation}
where the random variable $\chi_2$ has an $N$-independent distribution for large $N$,
${\rm Prob}(\chi_2\le x)= F_2(x)$ where $F_2(x)$ is the celebrated Tracy-Widom 
distribution~\cite{T-W mean} for $\beta=2$. However, for finite but large $N$, the
tails of the pdf of $\lambda_{\rm max}$ (for $|\lambda_{\rm max}-4N|\sim O(N)$) 
show significant deviations from the Tracy-Widom
behavior. The behavior in the tails of the pdf $P(\lambda_{\rm max}=t,N)$ 
is {\em instead} well described by
the following functional forms~\cite{Johansson}, valid for arbitrary $\beta$,
\begin{eqnarray}
P(t,N)&\sim & \exp\left[-\beta\, N^2\, \Phi_{-}
\left(\frac{4N-t}{N}\right)\right] \quad {\rm for} \,\, t\ll 4N \,;
 \label{lldv} \\
& \sim & \exp\left[-\beta\, N\, \Phi_{+}\left(\frac{t-4N}{N}\right)\right]
\quad {\rm for}\,\, t\gg 4 N \,;  
\label{rldv}
\end{eqnarray}
where $\Phi_{\pm}(x)$ are the right (left) large deviation (rate) functions for
the large positive (negative) fluctuations of $\lambda_{\rm max}$. Interestingly,
the explicit form of the rate functions $\Phi_{\pm}(x)$ have recently
been computed and were shown to be indepedent of $\beta$. 
The left rate function $\Phi_{-}(x)$ was computed 
in Ref.~\cite{vivo} using a Coulomb gas method developed in the context of
Gaussian random matrices~\cite{Dean} and is given for $x\ge 0$ by
\begin{equation}
\Phi_{-}(x)= \ln\left(\frac{2}{\sqrt{4-x}}\right)-\frac{x}{8}-\frac{x^2}{64}.
\label{lldv1}
\end{equation}
The right rate function $\Phi_{+}(x)$ was also computed very recently~\cite{vergassola}
using a different method 
\begin{equation}
\Phi_{+}(x)=
\frac{x+2}{2}-\ln(x+4)+\frac{1}{x+4}\,G
\left(\frac{4}{4+x}\right)\,,
\label{rldv1}
\end{equation}
where $G(z)=_3F_2\left[\{1,1,3/2\},\{2,3\}, z \right]$ is a
hypergeometric function. For small argument $x$, the two rate functions
have the following behavior~\cite{vivo,vergassola}
\begin{eqnarray}
\Phi_{-}(x) & \approx & x^3/{384} \label{lldv2} \\
\Phi_{+}(x) & \approx & x^{3/2}/6 \label{rldv2}
\end{eqnarray} 
Using these results, it was shown~\cite{vivo,vergassola} that both large deviation tails of the 
pdf of $\lambda_{\rm max}$
in Eqs. (\ref{lldv}) and (\ref{rldv}) match smoothly with the inner Tracy-Widom form. 

\begin{figure}[htp]
\begin{center}
\includegraphics[width=8.2cm]{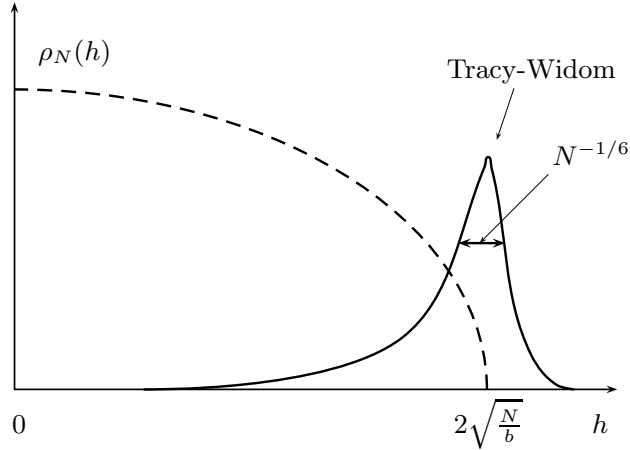}
\caption{The dashed line shows the density of states
$\rho_N(h)$
(quarter of ellipse).
The height of the upper interface (maximal height)
is centered around its mean $2\sqrt{\frac{N}{b}}$
with fluctuations of order $O(N^{-1/6})$
described by the Tracy-Widom law.
}\label{fig: TW}
\end{center}
\end{figure}

These results can then be directly translated
to our problem of interfaces identifying $b\, h_{\rm max}^2=\lambda_{\rm max}$.
For large $N$, the typical fluctuations of $h_{\rm max}$ around 
its mean are Tracy-Widom distributed (see figure \ref{fig: TW}) over 
a scale $\sim O(N^{-1/6})$. More precisely, we get
\begin{equation}
\label{TW}
h_{\rm max}\approx 2 \sqrt{\frac{N}{b}}+ 2^{-\frac{2}{3}} \,
 b^{-\frac{1}{2}} \: N^{-\frac{1}{6}}\, \chi_2
\end{equation}
where $\chi_2$ is Tracy-Widom distributed,
${\rm Prob}(\chi_2\le x)= F_2(x)$.

This gives in particular the first finite size correction
to the leading term for the average of the maximal height, in the large 
$N$ limit:
\begin{equation}
\langle  h_{\rm max} \rangle\approx  2 \sqrt{\frac{N}{b}}+ 2^{-\frac{2}{3}} \,
 b^{-\frac{1}{2}} \: N^{-\frac{1}{6}}\, \langle \chi_2 \rangle
\label{meanhmax}
\end{equation}
where $ \langle \chi_2 \rangle \approx -1.7711$~\cite{T-W mean}.
The variance can also be computed from \eqref{TW}
and the known variance of the Tracy-Widom distribution
 $\langle \chi_2^2 \rangle -\langle \chi_2 \rangle^2 \approx  0.8132$ 
\cite{T-W mean}:
\begin{eqnarray}
{\rm Var}(h_{\rm max})&=&\langle h_{\rm max}^2 \rangle-\langle h_{\rm max} \rangle^2 \nonumber \\
&\approx& \frac{2^{-4/3}}{b}
\, N^{-1/3}\left( \langle \chi_2^2 \rangle - \langle \chi_2 \rangle^2 \right)\nonumber \\
&\approx &\frac{0.32}{b \, N^{1/3}} 
\label{varhmax}
\end{eqnarray}

\hspace{-2cm}

Similarly, the {\em atypical large} fluctuations
of $h_{\rm max}$ around its mean (for $|h_{\rm max}-2\sqrt{N/b}|\sim O(N^{1/2})$)
are described by the large deviation tails as for $\lambda_{\rm max}$
in Eqs. (\ref{lldv}) and (\ref{rldv}), with $\beta=2$.
 For the left large deviation, we get
for $b t^2-4 N \sim O(N)$ with $bt^2 < 4 N$ :
\begin{equation}
{\rm P}\left[h_{\rm max}= t, N \right]
\approx 
 \exp\left\{-2 N^2\, \Phi_{-}
\left(\frac{4N-b t^2}{N}\right)\right\}
\label{hlldv1}
\end{equation}
where $\Phi_{-}(x)$ is given in Eq. (\ref{lldv1}).

Analogously the large and rare fluctuations to the right of the mean
can also be computed from the exact expression of $\Phi_{+}(x)$
in Eq. (\ref{rldv1}).
Replacing again $\lambda_{\rm max}$ by
$b\, h_{\rm max}^2$, we get the right large deviation tail
of the pdf of $h_{\rm max}$, for large $N$ and for
$b t^2-4 N \sim O(N)$ with $b t^2 > 4 N$:
\begin{equation}
{\rm P}\left[h_{\rm max}= t, N \right]
\approx 
 \exp\left\{-2 N\, \Phi_{+}
\left(\frac{b t^2-4 N}{N}\right)\right\}
\label{hrldv1}
\end{equation}
where $\Phi_+(x)$ is given in Eq. (\ref{rldv1}).

\begin{figure}[btp]
\centering
\begin{pspicture}(0,0)(6,6)
\centering
\includegraphics[width=7.5cm]{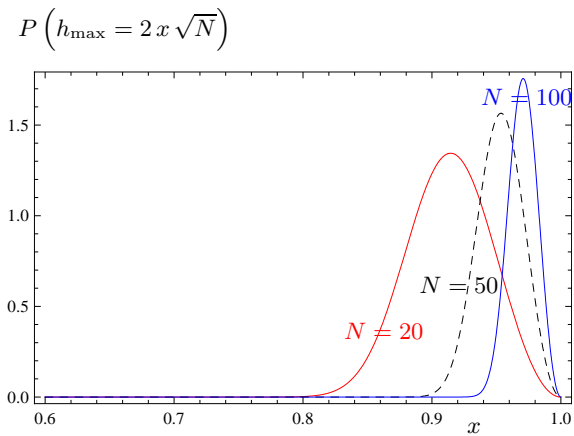}
\rput{0}(-6,5.3){ $P\left(h_{\rm max}= 2 \, x \, \sqrt{N}  \right)$}
\rput{0}(-1.3,0){$x$}
\rput{0}(-2.5,1.3){\red $N=20$}
\rput{0}(-1.5,1.9){\black $N=50$}
\rput{0}(-.6,4.4){\blue $N=100$}
\end{pspicture}
\caption{Pdf of the maximum of the heights (large deviation),
plotted for $b=1$ and for different $N$, as a function
of the rescaled height $x=\frac{t}{2 \, \sqrt{N}}$. As $N$ increases,
the rescaled location of the peak of the pdf approaches to $x\to 1$.
The width of the regime around the peak where the Tracy-Widom law is valid 
is reduced to $\sim N^{-1/6-1/2}\sim N^{-2/3}$ in this scale. The
rest of the pdf beyond the peak are described by the large deviation tails 
in Eqs. (\ref{hlldv1}) and (\ref{hrldv1}).}
\end{figure}

\subsection{Minimal height of the interfaces}
\label{subsec:min}

We have seen in subsection \ref{subsec:dens} that the lower bound of the 
support of the average
density of interfaces
is indeed zero in the first approximation as $N\to \infty$.
Since the lower edge of the support is also precisely
the average height of the lowest (close to the substrate) interface,
we have $\langle h_{\rm min}\rangle \to 0$ as $N\to \infty$.
This is clearly an effect of the fermionic repulsion between the
interfaces, because for `independent' interfaces (that are allowed to cross)
the height of the lowest interface (minimal height)
is of order $h_{\rm min}\sim O(1)$:
it does not see the other interfaces.
The hard wall at the origin in the problem of interfaces
corresponds to the constraint that the Wishart eigenvalues
must be positive. In the Wishart ensemble, the neighborhood
of the origin is, thus, called the hard edge of the spectrum.
The distribution for the hard edge is known to be related to the 
Bessel kernel \cite{Forrester,TWbis}, just like the distribution
at the soft edge (for the maximum eigenvalue) is related to the Airy kernel.

To know more precisely how $h_{\rm min}$ decreases with increasing $N$
in presence of the `nonintersection' constraint, we need
to find the statistics of $h_{\rm min}$ for large but {\em finite} $N$, which
is precisely the objective of this subsection.
The main result of this subsection is
to show that for nonintersecting interfaces,
the minimal height is typically of order $h_{\rm min}\sim O\left(\frac{1}{\sqrt{N}}\right)$
to leading order in large $N$. For special values of the parameters,
we are also able to calculate the {\em full} distribution
of the minimal height $h_{\rm min}$ as discussed below.

Under the customary change of variables $bh_i^2=\lambda_i$, it follows that
$h_{\rm min}$ has the same distribution as $\sqrt{\lambda_{\rm min}/b}$
where $\lambda_{\rm min}$ is the minimum eigenvalue of Wishart ensemble
with parameters $\beta=2$ and arbitrary $M\ge N$. The minimum eigenvalue
of the Wishart ensemble has been studied before~\cite{edelman}, with
applications in the quantum entanglement problem in bipartite systems~\cite{majumdar-quant}. 
When $M-N\sim O(1)$, which is precisely our case since $\alpha=M-N+1/2\sim O(1)$,
the minimum eigenvalue is known to scale, for large $N$, as $\lambda_{\rm min}\sim 1/N$
for arbitrary $\beta$, though it has only been proved exactly for special
values of $M-N$ and $\beta$, e.g., for $\beta=1$ and $M=N$ or for $\beta=2$ and $M=N$~\cite{edelman}.   
For the interface model it then follows quite generally that $h_{\rm min}\sim 1/\sqrt{N}$ for large 
$N$ for general $\alpha \sim O(1)$.
The fermionic repulsion between interfaces has
thus again a strong effect: 
the lowest interface is pushed very close to the substrate since 
its height is of order $h_{\rm min}\sim 1/\sqrt{N}$ 
for large $N$ (instead of $O(1)$ for non-interacting interfaces that can cross).
 
To go beyond this scaling behavior for large $N$ and compute precisely the
statistics of $h_{\rm min}$ for arbitrary $N$ seems difficult for general $\alpha$. Below
we show that for the special case $\alpha=3/2$, it is possible
to compute the full distribution of $h_{\rm min}$ for all $N$.  

It turns out to be convenient to compute
the cumulative distribution function (cdf) of
the minimal height ${\rm Prob}\left[h_{\rm min}\ge \sqrt{\zeta}, N\right]$, for arbitrary
$\zeta$. We use the notation $\sqrt{\zeta}$ for the convenience of scaling
as seen below. Our starting point is the central result for the 
joint pdf of interface heights in Eq. (\ref{jpdfh1}). Clearly, the
event that the minimum height $h_{\rm min} \ge \sqrt{\zeta}$ is
equivalent to the event that all the heights are greater than $\sqrt{\zeta}$:
$h_i\ge \sqrt{\zeta}$ for all $i=1,2,\ldots,N$. 
Hence,
\begin{equation}
{\rm Prob}\left[h_{\rm min}\ge \sqrt{\zeta}, N\right]= 
\int_{\sqrt{\zeta}}^{\infty}dh_1\ldots \int_{\sqrt{\zeta}}^{\infty} dh_N\, P(h_1,h_2,\ldots,h_N)
\label{minh1}
\end{equation}
where the joint pdf $P(h_1,h_2,\ldots, h_N)$ is given in Eq. (\ref{jpdfh1}).
Making the standard change of variables, $bh_i^2=\lambda_i$ 
we then have 
\begin{equation}
{\rm Prob}\left[h_{\rm min}\ge \sqrt{\zeta}, N \right] 
= A_N \int_{b\zeta}^{\infty} d\lambda_1\ldots \int_{b\zeta}^{\infty} d\lambda_N \:
 e^{-   \sum_k \lambda_k } 
\; \prod_k \lambda_k^{\alpha-\frac{1}{2}} \;\:
\prod_{i<j} (\lambda_i-\lambda_j)^2  
\label{cdfmin1}
\end{equation}
where the normalization constant $A_N$ is given in Eq. (\ref{normal1}).
Next, making a shift $\lambda_i=b\zeta+x_i$, one can rewrite Eq. (\ref{cdfmin1})
in a more compact form
\begin{equation}
{\rm Prob}\left[h_{\rm min}\ge \sqrt{\zeta},N \right]= A_N\, e^{-b N \zeta}\, w(b \zeta)
\label{cdfmin2}
\end{equation}
where the function $w(z)$ is given by the multiple integral
\begin{equation}
w(z)=\int_0^{\infty}dx_1\ldots \int_0^{\infty}dx_N \;
 e^{-   \sum_k x_k } 
\; \prod_{k=1}^N (x_k+z)^{\alpha-\frac{1}{2}} \;\:
\prod_{i<j} (x_i-x_j)^2.
\label{wz1}
\end{equation}
For notational simplicity we have suppressed the $N$ and $\alpha$ dependence 
of $w(z)$.
The multiple integral in Eq. (\ref{wz1}) is not easy to evaluate for
general values of the parameter $\alpha$.  
However, one can make progress for  special values of $\alpha$. 

We first note that when the parameter $\alpha-\frac{1}{2}=M-N$ is an integer,
$ w(z)$ is a polynomial of $z$ of degree $\alpha-\frac{1}{2}$.
In the special case $\alpha=3/2$, i.e., $M=N+1$, one can
explicitly evaluate $w(z)$ by following a method similar
to the one used by Edelman~\cite{edelman} to compute the
distribution of $\lambda_{\rm min}$ for Wishart matrices with 
$\beta=1$ and $M=N$. In this special case
$M=N+1$ and $\beta=2$, the distribution of the minimum eigenvalue,
in the large-$N$ limit, was already computed by Forrester \cite{Forrester}.
However, in this special case, one can actually calculate the distribution
and the moments even for all finite $N$ as we demonstrate below.
For $\alpha=3/2$ and $\beta=2$, 
we first compute two derivatives $w'(z)$ and $w''(z)$
of the function $w(z)$ in Eq. (\ref{wz1}). Using integration by parts
and some rearrangements, we find that 
$w(z)$ satisfies an ordinary second order differential equation for
any $N$
\begin{equation}
z \, w''(z) + (1+z ) \,w'(z)-N\, w(z)=0
\label{wdiff1}
\end{equation}
whose unique (up to a constant) solution is in fact
the ordinary Laguerre polynomial with negative argument
\begin{equation}
w(z) \propto \mathcal{L}_N (-z)
=\sum_{k=0}^N \left(\begin{array}{c} N\\ k\end{array} \right)
 \frac{z^k}{k!}
\label{wsol1} 
\end{equation}
and finally, since ${\rm Prob}\left[h_{\rm min}\geq 0, N \right]=1$, we get
\begin{equation}
\label{eq:min}
{\rm Prob}\left[h_{\rm min}\geq t, N  \right]
=e^{-b N t^2} \: \mathcal{L}_{N}(-b\, t^2)
=e^{-b N t^2} \: \sum_{k=0}^N \left(\begin{array}{c} N\\ k\end{array}
\right)
 \, \frac{b^k \, t^{2 k}}{k!}. 
\end{equation}
Hence, for the special case $\alpha=3/2$ we can then give 
an explicit expression
for the pdf of $h_{\rm min}$ valid for all $N$,
\begin{equation}
\label{eq:minPDF}
P\left(h_{\rm min}= t, N \right)
=-\frac{d}{dt} {\rm Prob}\left[h_{\rm min}\geq t, N  \right] 
=2 \,b^2 \,  t^3 \,  e^{-b N t^2} \:
\mathcal{L}_{N-1}^{(2)}(-b\, t^2)
\end{equation}
where $\mathcal{L}_n^{\gamma}(x)$ is the generalized Laguerre polynomial
already defined in Eq. (\ref{laguerre1}).

From the exact pdf in Eq. (\ref{eq:minPDF}) one can calculate
all its moments explicitly as well (see appendix-A for details). We find
for the $k$-th moment, for arbitrary $N$,
\begin{equation}
\langle h_{\rm min}^k\rangle= 
\frac{\Gamma(k/2+2)}{2b^{k/2}}\,\frac{(N+1)}{N^{k/2+1}}\,_2F_1\left(-(N-1),k/2+2;3;-1/N\right).
\label{momhmin1}
\end{equation}
One can then work out the asymptotic behavior of the moments for large $N$.
For example, one can show (see appendix-A)
that the average value ($k=1$) $\langle h_{\rm min}\rangle$
decays for large $N$ as
\begin{equation}
\langle h_{\rm min}\rangle \approx \frac{c_1}{\sqrt{b N}}
\label{avghmin}
\end{equation}
with the constant prefactor $c_1$ given exactly by
\begin{equation} 
c_1= \sqrt{\frac{\pi e}{4}}\, I_0(1/2)= 1.5538\ldots
\label{prefac1}
\end{equation}
where $I_n(z)$ is the modified Bessel function of the first kind with index $n$
\begin{equation}
I_n(z)= \sum_{k=0}^{\infty} \frac{1}{k! \, (k+n)!}
\left(\frac{z}{2}\right)^{2 k +n}.
\label{modbessel}
\end{equation}
 
For large $N$, one can also work out precisely the scaling behavior
of the full pdf of $h_{\rm min}$ given in Eq. (\ref{eq:minPDF})
and recover the result of Forrester \cite{Forrester}.
Since typically $h_{\rm min}\sim 1/\sqrt{N}$, one expects that
its pdf (normalized to unity) has a scaling form for large $N$ 
\begin{equation}
P\left(h_{\rm min}= t, N  \right)\approx \sqrt{N} \, f_{\rm min}(t \, \sqrt{N}).
\end{equation} 
 The scaling function $f_{\rm min}(x)$ can be computed explicitly
from Eq. (\ref{eq:minPDF}). We get
\begin{eqnarray}
&&f_{\rm min}(x)=\lim_{N \rightarrow \infty} \frac{1}{\sqrt{N}}
P\left(h_{\rm min}= \frac{x}{\sqrt{N}}\, ,N  \right) \nonumber\\
&&\;\;\; \;\; =\lim_{N \rightarrow \infty} 
b^2 \, x^3  e^{-b \, x^2}\:  \,_1F_1(1-N ; 3 ;\frac{b
  x^2}{N})
\nonumber \\
&&\;\;\; \;\; =\: b^2\,x^3\,e^{-b\, x^2}\, \,_0F_1(3 ; b\, x^2) \nonumber\\
&&\;\;\; \;\; =\: 2 b \, x \, e^{-b\, x^2}\: I_2 (2 x \sqrt{b}). 
\label{scalingfmin}
\end{eqnarray}
This function has the following asymptotic behavior
\begin{eqnarray}
f_{\rm min}(x) & \approx & b^2\, x^3 \quad {\rm as} \,\, x\to 0 \nonumber \\
& \approx & \frac{b^{3/4}}{\sqrt{\pi}}\, \sqrt{x}\, e^{-b\, x^2 + 2\,\sqrt{b}\,x} \quad {\rm as} \,\, 
x\to \infty
\label{fminasymp}
\end{eqnarray}
A plot of this scaling function is given in Fig. 6.

\begin{figure}[btp]
\centering
\begin{pspicture}(0,0)(6,6)
\centering
\includegraphics[width=7.8cm]{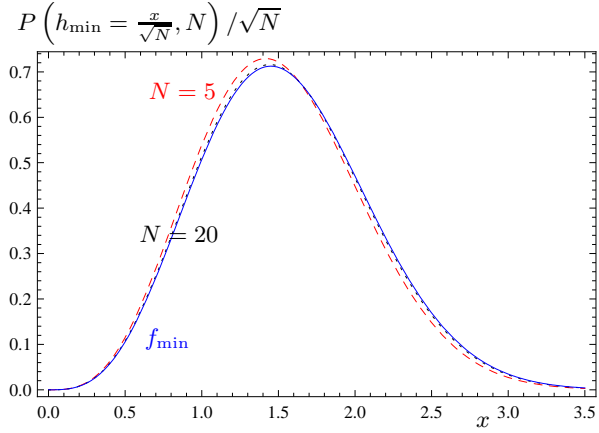}
\rput{0}(-6,5.3){ $P\left(h_{\rm min}= \frac{x}{\sqrt{N}}, N  \right)/\sqrt{N}$}
\rput{0}(-1.5,0){$x$}
\rput{0}(-5.5,4.4){\red $N=5$}
\rput{0}(-5.6,2.5){ $N=20$}
\rput{0}(-5.7,1.1){\blue $f_{\rm min}$}
\end{pspicture}
\caption{Rescaled probability density function of the minimum of the heights,
in the case $\alpha=3/2$,
plotted for $b=1$ and for $N=5$ (dashed line) and $N=20$ (dotted line)
 and compared
with the limiting distribution $f_{\rm min}(x)$ (solid line). 
In the large $N$ limit, 
indeed the curves approach the limiting distribution $f_{\rm min}(x)$.}
\end{figure}

\subsection{Center of mass}
\label{subsec:cdm}

We study in this subsection the distribution of the center of mass
of the heights
\begin{equation}
\label{cdmDef}
G_N=\frac{h_1+...+h_N}{N}
\end{equation}
for large $N$. 
If the interfaces were allowed to cross,
the heights $h_i$ would be independent and
identically distributed variables (i.i.d.).
In that case, the distribution of the center of mass $G_N$ would
 be a pure Gaussian distribution
in the large $N$ limit (central limit theorem).
But in our model, due to the repulsion between the interfaces,
 the interface heights are strongly correlated.
What is the effect of the repulsion 
(nonintersecting constraint) on the center of mass?
We will show how to compute
the pdf of the center of mass for large $N$ upon
borrowing some techniques developed in the context of random matrix theory,
using in particular the analogy between the  Wishart eigenvalues
and  a Coulomb gas of charges.
We will show that the pdf of the center
of mass $P(G_N=\nu)$ has an extraordinarily weak non-analytic behavior
at $\nu=\langle G_N\rangle$ (where $\langle G_N\rangle$ is
the average of the center of mass), which is shown to be 
a direct consequence of a phase
transition in the associated Coulomb gas problem.

Since the typical height of an interface 
$h_{\rm typ} \sim \sqrt{N}$ for large $N$, it follows
that the center of mass $G_N \approx O(\sqrt{N})$.
More precisely, by symmetry of the joint pdf of the heights,
the average of the center of  mass
is given by the average height (see \eqref{meanh}):
\begin{equation}
\label{cdm:mean}
\langle G_N \rangle =\langle h \rangle
\approx \frac{8}{3 \pi }\,  \sqrt{\frac{N}{b}}
\equiv \mu \sqrt{N}
\;\; \textrm{for large $N$,}\;\;
\textrm{where}\;\; \mu= \frac{8}{3 \pi \sqrt{b}}
\end{equation}
Let us thus write $\nu=s \sqrt{N}$, where the scaled variable $s\sim O(1)$.

The main result of this subsection is to show that 
in the scaling limit $N\to \infty$, $\nu\to \infty$
but keeping the ratio $s= \nu/\sqrt{N}$ fixed, the 
pdf of the center of mass scales as:
\begin{equation}
P\left(G_N=\nu \right)\propto \exp\left[- N^2 \: 
\Phi\left(\frac{\nu}{\sqrt{N}}\right)\right]
\label{ldvcm}
\end{equation}
where the associated large deviation function $\Phi(s)$
 is plotted in figure \ref{fig:phicdm},
 has the following asymptotic behavior
\begin{eqnarray}
\Phi(s)\approx 
\left\{\begin{array}{lcl}
-2\ln s \;\;\; &{\rm for} &\;\; s\rightarrow 0^+ \\ && \\
\;\;\; b\, s^2 \;\;\;\; &{\rm for} &\;\; s\rightarrow +\infty 
 \end{array}\right. 
\end{eqnarray}
and  is a non-analytic smooth function:
$\Phi(s)$ is
infinitely differentiable everywhere but it is not analytic.
More precisely, we will show that $\Phi(s)$ is given by
\begin{equation}
\Phi(s)=\left\{\begin{array}{l}
\Phi^-(s)\;\; {\rm for} \;\; s<\mu \\
\Phi^+(s)\;\; {\rm for} \;\; s>\mu
\end{array} \right.
\;\; \;\; \textrm{where $\Phi^-$
and $\Phi^+$ are analytic functions
on their domain of definition}
\footnote[1]{  $\Phi^-$
 is analytic on the complex plane except for a branch cut 
along the negative real axis; and  $\Phi^+$ is analytic
on an open set of the complex plane including the (real) half-line
$s>\mu$.}
\end{equation}
\vspace{-.3cm}
\begin{eqnarray}
\textrm{and  with} \;\;\;\;\;\;\;\;
\Phi^+(s) - \Phi^-(s)\approx
-\pi \, \sqrt{b}\,
(s-\mu)\,
 e^{-\frac{8}{\pi \, \sqrt{b} \, (s-\mu) }}\: e^{4 (\ln 2-1)}\: 
 \;\;\;\;\;
\;\;{\rm as }\;\; s\rightarrow \mu^+
\end{eqnarray}
$\Phi(s)$ has thus an essential singularity at $s=\mu$, it is not analytic.

But all the derivatives of $\Phi$ exist and are continuous.
In particular, for $s\rightarrow\mu$,
$\Phi$ has, in first approximation, a quadratic behavior:
\begin{equation}
\Phi(s)\approx \frac{(s-\mu)^2}{2\,\sigma^2} \;\;\;
 {\rm for} \;\; s\rightarrow \mu\
\;\; {\rm with} \;
 \; \mu=\frac{8}{3 \, \pi\,\sqrt{b}}
\;\; {\rm and} \;\; \sigma=\frac{1}{\pi}\sqrt{\frac{2}{b}}
\end{equation}
The pdf of the center of mass can thus be approximated by a Gaussian around
its minimum ($s=\mu$), which
 gives the mean and variance of the center of mass:
\begin{eqnarray}
\langle G_N \rangle\approx\mu \sqrt{N}
 \approx \frac{8}{3 \pi} \sqrt{\frac{N}{b}}\;\;\;{\rm and}\;\;\;
\sqrt{\textrm{Var}(G_N)}=
\sqrt{\langle G_N^2\rangle-\langle G_N\rangle^2}
  \approx \frac{\sigma}{\sqrt{N}}
\approx 
\frac{1}{\pi}\sqrt{\frac{2}{N\, b}}
\end{eqnarray}
We will also derive an exact closed form for $\Phi^-(s)$: 
\begin{equation}
\Phi^-(s)=\frac{L(s)^2}{32}\,-\,\ln\left(\frac{L(s)}{4}\right)-\frac{1}{2}
\;\;
{\rm where}\;\;
L(s)=\left[ 2^{5/3}\, g_1(s)^{-1/3}- 
2^{1/3}\, g_1(s)^{1/3} \right]^2 \;\; {\rm with} \;\; g_1(s)=
-3\pi s\,\sqrt{b}+\sqrt{16+9\, \pi^2 \, b\, s^2}
\end{equation}
But we will see that $\Phi^+(s)$ is more difficult to compute:
we will only derive its asymptotics ($s\rightarrow +\infty$ and
$s\rightarrow\mu^+$).
\\
\\
\\

To derive these results, let us start with 
the pdf of the center of mass:
\begin{eqnarray}
\label{cdmPdf}
\hspace{-.5cm} P(G_N=s\,\sqrt{N})&=&
\int_0^{\infty} dh_1 ...\int_0^{\infty} dh_N  \;
 \delta\left(\frac{h_1+...+h_N}{N}-s\,\sqrt{N} \right) \:
 P(h_1,...,h_N) \nonumber \\
&=& A_N \: \int_0^{\infty} d\lambda_1 ... \int_0^{\infty}
 d\lambda_N \;  e^{-   \sum_k \lambda_k } 
\; \prod_k \lambda_k^{ \alpha- \frac{1}{2}} \;\:
\prod_{i<j} (\lambda_i-\lambda_j)^2 \; \delta
 \left(\frac{ \sqrt{\lambda_1}+...
+\sqrt{\lambda_N}}{N\,\sqrt{b} }-s\,\sqrt{N} \right)
\end{eqnarray}
where we have used Eq. 
(\ref{jpdfh2}). The integrand (without the delta function)
can be written as $\exp\left[-E\{\lambda_i\}\right]$ where
\begin{equation}
E\{\lambda_i\}= \sum_{i=1}^N \lambda_i- \left(\alpha-\frac{1}{2}\right)\sum_{i=1}^N \ln (\lambda_i)
-\sum_{j\ne k} \ln |\lambda_j-\lambda_k|
\label{energyC}
\end{equation}
can be interpreted as the energy of a Coulomb gas of $N$ charges with coordinates $\{\lambda_i\}$
as mentioned earlier in section III.
Thus the calculation of the distribution of the center of mass reduces to the calculation of the 
distribution of a particular functional of this Coulomb gas. This can be performed exactly for large $N$
using a functional integral method followed by saddle point calculations.
This method has been used recently in several contexts: for example, to calculate
the large fluctuations of the maximum eigenvalue of both Gaussian and Wishart random matrices
~\cite{Dean,vivo,vergassola}, to compute the purity distribution in bipartite entanglement
of a random pure state~\cite{facchi} and also to compute the distributions 
of conductance and shot noise for ballistic transport in a chaotic cavity~\cite{shotnoise}. 

To evaluate the multiple integral in Eq. (\ref{cdmPdf}) by the functional integral
method one proceeds in two steps.
First step is a coarse-graining procedure that sums over (partial tracing)
all microscopic configurations of $\{\lambda_i\}$ compatible
with a fixed normalized (to unity) charge density $\rho_N(\lambda)= N^{-1}\sum_{i} 
\delta(\lambda-\lambda_i)$ 
and a fixed value of $s= \sum_i \sqrt{\lambda_i}/\left({N^{3/2}\sqrt{b}}\right)$. The next step
is to integrate over all possible normalized charge densities with fixed $s$--this
is the functional integration which is then carried out using the method of steepest
descent for large $N$. 

To proceed, we first scale the positions of charges,
$x=\frac{\lambda}{N}$ such that $x\sim O(1)$ and define the charge density in the $x$ space 
$\rho(x)=\frac{1}{N} \sum_i \delta\left(x-\frac{\lambda_i}{N} \right)$.
With this scaling, it is easy to check that while the first and the third term
in the energy expression in Eq. (\ref{energyC}) are both of order $\sim O(N^2)$, the
second term multiplying $(\alpha-1/2)$ (corresponding to the external logarithmic potential)
is of order $\sim O(N)$, as long as $(\alpha-1/2)$ is of order $\sim O(1)$. Thus this
term becomes negligible for large $N$ for any $N$-independent $\alpha$ and to leading order
in large $N$, the $\alpha$-dependence just drops out.
Then, the coarse-graining procedure gives, to leading order for large $N$,
\begin{equation}
\label{functInt}
P(G_N=s \sqrt{N}) \propto
\int \mathcal{D}\left[ \rho \right] \, e^{-N^2 \, E_s\left[ \rho \right]+O(N)}
\end{equation}
where the effective energy functional is given by:
\begin{eqnarray}
\label{cdmEffEner}
 \hspace{-.5cm} E_s\left[ \rho \right]= \int_0^{\infty} x \rho(x) dx 
 -\int_0^{\infty}
 \int_0^{\infty} \rho(x)\rho(x')
\ln|x-x'| dx dx' \:+R  \left( \int_0^{\infty} \sqrt{x}
 \rho(x) dx -s\,\sqrt{b} \right)
  + D \left( \int_0^{\infty} \rho(x)dx -1 \right)
\end{eqnarray} 
We have introduced two Lagrange multipliers, $R$ and $D$,
in order to take into account two constraints.
The first (associated to $R$) enforces 
the condition
$G_N=\frac{\sqrt{\lambda_1}+...+\sqrt{\lambda_N}}{N\, \sqrt{b}}=s \sqrt{N}$
or equivalently $ \int_0^{\infty} \sqrt{x}
 \rho(x) dx = s\,\sqrt{b}$
(it replaces the delta function in the expression
of $P\left(G_N=s \sqrt{N} \right)$).
The second (associated to $D$)
enforces the normalization of the density $\rho$:
$\int_0^{\infty} \rho(x)dx =1$.

The functional integral \eqref{functInt} is carried out in the large 
$N$ limit by  the method of steepest descent. Hence:  
\begin{equation}
P(G_N=s \sqrt{N}) \propto \exp\left[-N^2 \, E_s\left[ \rho_c \right]\right]
\end{equation}
where $\rho_c(x)$
 minimizes the effective energy: $\frac{\delta E_s[\rho(x)]}{\delta \rho(x)}=0$.
The saddle point density $\rho_c(x)$ 
is thus given by the equation:
\begin{equation}
\label{cdm:SaddlePoint}
x + R \sqrt{x} + D=2 \int_0^{\infty} \rho_c(x') \ln|x-x'| dx'
\end{equation}
Differentiating once with respect to $x$ leads to the integral equation:
\begin{equation}
\label{cdm:hilbert}
1 +\frac{R}{2 \sqrt{x}}=
2 \mathcal{P} \int_0^{\infty} \frac{\rho_c(x')}{x-x'} dx'=2 \, H_x\left[
 \rho_c\right]
\end{equation}
 $H_x\left[ \rho_c\right]$ is called the semi-infinite
Hilbert transform of $\rho_c$ (and $\mathcal{P}$
denotes the principal value). It
is not easy to invert it directly.
However, the finite Hilbert transform
$ H_x^f\left[y\right]=\mathcal{P}\int_a^b \frac{y(t)}{t-x}dt$
can be inverted using a theorem proved by Tricomi \cite{tricomi}.
According to Tricomi, the solution of the integral equation
\begin{equation}
\label{cdmTricomi1}
f(x)=\mathcal{P}\int_a^b \frac{y(t)}{t-x}dt \;\;
\textrm{with}\;\; a<x<b ,\; |a|+|b|<\infty
\end{equation}
is given by
\begin{eqnarray}
\label{cdmTricomi2}
 y(x)=\frac{1}{\pi^2 \sqrt{x-a}\,\sqrt{b-x}} 
\left[C_0-\mathcal{P}\int_a^b \frac{\sqrt{t-a }\,\sqrt{b-t}}{t-x} f(t)dt \right]
\;\;\;\;\;
 \textrm{for} \;\; a<x<b
\;\;\;\;\;\;\;\;\;\;\;\;\;\;\;\;\;\;\;\;\;\;\;
\end{eqnarray}
where $C_0$ is an arbitrary constant.
Tricomi showed that $C_0$ then satisfies:
$ \pi \int_a^b y(t)dt=C_0$.
We will hereafter assume that the saddle point density $\rho_c$
has a finite support and use Tricomi's result. 

So, the steps we need to carry out are (i) to find the solution $\rho_c(x)$ of the integral
equation (\ref{cdm:hilbert}) which will contain yet unknown Lagrange multipliers $R$ and $D$
(ii) fix $R$ and $D$
 from the two conditions: $\int_0^{\infty} \rho_c(x)\, dx=1$ and
$\int_0^{\infty} \sqrt{x}\, \rho_c(x)\, dx = s\sqrt{b}$ for a fixed given $s$ and
(iii) evaluate the saddle point energy $E_s[\rho_c]$ which is then
precisely
(up to an additive constant)
the large deviation function $\Phi(s)$ announced in Eq. (\ref{ldvcm}). 

Physically, as the effective (external) potential for the charges is
of the form $V_f(x)= x +R \sqrt{x}+D$
(see equations \eqref{cdmEffEner}
and \eqref{cdm:SaddlePoint}), we expect a 
different behavior of the charge density $\rho_c(x)$ depending on the sign
of the Lagrange multiplier $R$.
\vspace{0.3cm}

\noindent  $\bullet$ For $R>0$, the effective potential
$V_f(x)= x +R \sqrt{x}+D$
is an increasing function of
$x$ for $x\geq 0$ with minimum at $x=0$.
In this case, the charges will be confined near the origin.
Therefore the density must be large for small
$x$, decreasing as $x$ increases and finally vanishing at a certain $x=L$.
We thus assume that $\rho_c(x)$ has a finite support over $]0,L]$
where $L$ is fixed by demanding that the density vanishes at $x=L$:
 $\rho_c(L)=0$.
\vspace{0.3cm}

\noindent $\bullet$ However, for $R<0$, the effective potential
is minimal for
$x=x_0=\frac{R^2}{4 } >0$.
The density must be larger around $x=x_0$.
In that case, 
$\rho_c(x)$ will have a finite support over $[L_1,L_2]$
with $L_1 >0$ and where $L_1$ and $L_2$ are 
fixed by the constraints $\rho_c(L_1)=0=\rho_c(L_2)$.
\vspace{0.3cm}

We will see later that $R>0$ corresponds to the left side
of the mean of the center of mass ($s<\mu$), and $R<0$ corresponds
to its right side ($s>\mu$).
Thus there is a phase transition in this Coulomb gas problem as one tunes 
$s$ through $s=\mu$ or equivalently $R$ through the critical value $R=0$. The optimal charge density has 
different 
behaviors
for $R>0$ and $R<0$. When expressed as a function of $s$, this 
leads to non-analytic behavior of the saddle point energy, i.e., the
large deviation function $\Phi(s)$ at its minimum $s=\mu$. 

\subsubsection{Case $R \geq 0$ ($s \leq \mu$)}

Let us begin with the case $R \geq 0$, that will be shown
to correspond to
$s\leq \mu$ (left side of the mean of the center of mass).
In this case,
 the effective potential 
  is minimal for $x=0$.
We can thus assume that $\rho_c$ has a finite support over $]0,L]$
where $L$ is fixed by the constraint $\rho_c(L)=0$.
In this subsection, we compute the saddle point density $\rho_c(x)$
and derive an exact closed form for the energy $E_s[\rho_c]$
(and thus the function $\Phi(s)$). From this explicit form, we work out 
the asymptotic behavior of $\Phi(s)$ for $s\rightarrow 0$ and  for
$s\rightarrow\mu^-$. For $s\rightarrow\mu^-$, we will see
that the pdf can be approximated by a Gaussian -and this will give
 the mean and variance
of the pdf of the center of mass.
\\

The (normalized) solution $\rho_c(x)$, with support over $]0,L]$, 
of the integral equation \eqref{cdm:hilbert} can then be obtained
using Tricomi's theorem in Eq. (\ref{cdmTricomi2}). The resulting integral
can be performed using the Mathematica and we get
\begin{eqnarray}
\label{cdmR+Dens}
\rho_c(x)=\frac{1}{2 \pi} \sqrt{\frac{L-x}{x}}
+\frac{R}{2 \pi^2 \sqrt{x}}\: \argtanh \left( \sqrt{1-\frac{x}{L}} \right)
\;\;\;\;\textrm{for}\;\; 0<x\leq L
\end{eqnarray} 
where $\argtanh$ is the inverse hyperbolic tangent.

\begin{figure}[htp]
\centering
\begin{pspicture}(0,0)(6,6)
\includegraphics[width=7.8cm]{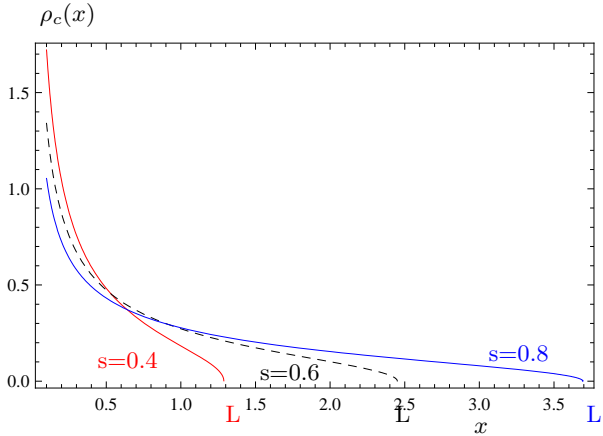}
\rput{0}(-1,.8){\blue s=0.8}
\rput{0}(0,0){\blue L}
\rput{0}(-4.1,0.55){ s=0.6}
\rput{0}(-2.6,0){ L}
\rput{0}(-6.2,.7){\red s=0.4}
\rput{0}(-4.8,0){\red L}
\rput{0}(-7,5.3){$\rho_c(x)$}
\rput{0}(-1.5,-0.2){$x$}
\end{pspicture}
\caption[]{
Density of states $\rho_c(x)$ (density of charges)
of the Coulomb gas associated to the computation of the pdf 
$P(G_N=s \sqrt{N})$ of the center of mass, 
in the case $s\leq \mu=\frac{8}{3 \pi \sqrt{b}}$ ($R\geq 0$),
plotted for different values of $s$ (and for $b=1$).
 The effective potential seen by the charges is minimal for $x=0$,
thus the density has a finite
support over $]0,L]$ and diverges at the origin.
\\
  $\bullet$ When $s$ tends to
  $\mu\approx 0.85$ for $b=1$
 (i.e. the center of mass tends to its mean value),
 $L$ tends to $4$ and $\rho_c$ tends to the average value of
the density of states ($R\rightarrow 0$).
\\
$\bullet$ When $s<\mu$ and $s$ decreases (i.e. the center
of mass is smaller than its mean and decreases),
$L<4$ and $L$ decreases also: the Coulomb gas of charges
 is  more and more compressed, the charges are more and more confined
close to the origin.}
\label{fig:dens R>0}
\end{figure}

As the density $\rho_c(x)$ must be positive for all $x \in ]0,L[$
(it is a density of states, of charges), such a solution
(with support over $]0,L]$) can exist only for $R\geq 0$.
For $R \neq 0$, we have indeed $\rho_c(x)\approx
\frac{R}{4 \pi^2}\:\frac{|\ln x|}{ \sqrt{x}}$ as $x\rightarrow 0^+$.
Therefore $R$ must be positive: $R\geq 0$. Conversely,
 it is not difficult to see 
that for $R\geq 0$, the density given
in Eq. \eqref{cdmR+Dens} is positive for all $x\in]0,L[$.
In this phase ($R \geq 0$), as figure \ref{fig:dens R>0}  shows,
the Coulomb charges
are confined close to the origin:
the interfaces are bound to the substrate.

We want to compute the pdf $P(G_N=s \,\sqrt{N})$. The basic variable
is thus $s$. There are also three unknown parameters: $R$ and $D$ are two
Lagrange
multipliers and $L$  is the upper bound of the density support.
These parameters will be determined  by enforcing the three constraints
 $\int_0^{\infty} \rho_c(x)dx =1$,  $ \int_0^{\infty} \sqrt{x}
 \rho_c(x) dx = s\,\sqrt{b}$ and $\rho_c(L)=0$. 

\begin{figure}[htp]
\centering
\begin{pspicture}(0,0)(6,4)
\includegraphics[width=7.5cm]{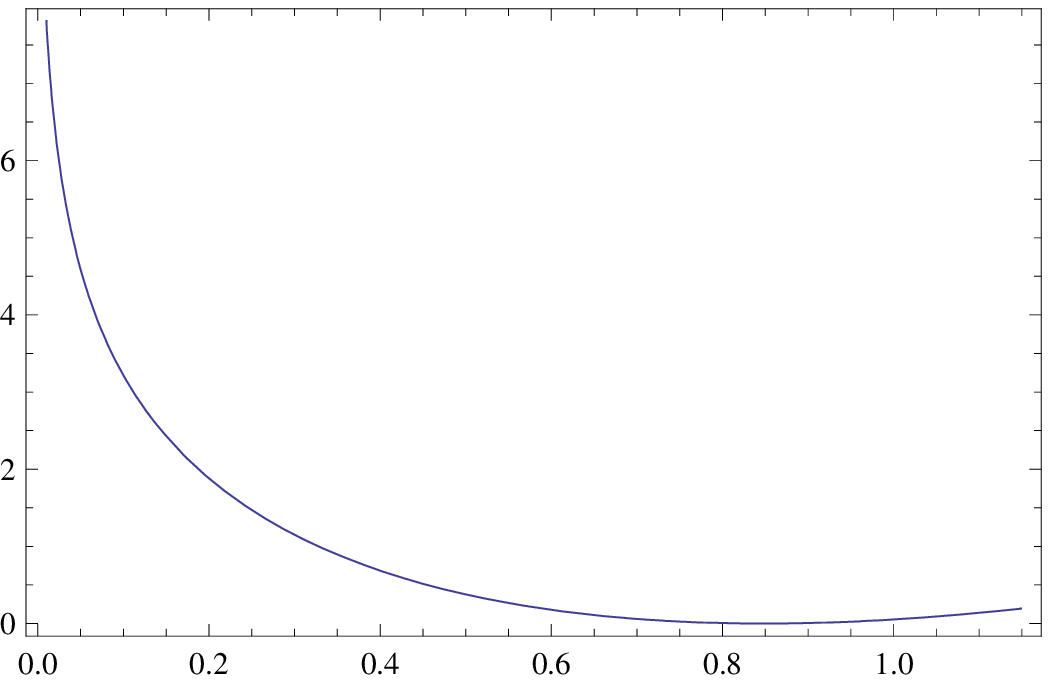}
\psline[linewidth=1pt,linecolor=black](-2,0.1)(-2,0.3) 
\rput{0}(-2,-.2){ $\mu$}
\rput{0}(-8,4.5){$\Phi(s)$}
\rput{0}(-.6,-0.2){$s$}
\end{pspicture}
\caption{Large deviation function $\Phi(s)$
of the pdf of the center of mass,
such that $P(G_N=s \sqrt{N})\propto e^{-N^2 \Phi(s)}$.
The minimum of $\Phi(s)$ occurs at $s=\mu= 8/\left(3 \pi \sqrt{b}\right)$
which corresponds to the average value of the center of mass.
We have chosen $b=1$ so that $\mu= 8/\left(3\pi\right)= 0.848826..$.
The domain $s<\mu$ corresponds to $R>0$ where the explicit form
of $\Phi(s) $ is known (Eq. \eqref{cdm:pdf}).
$\Phi(s)$ is a smooth function with a very weak non-analyticity
at $s=\mu$ (essential singularity) -that can not be seen
in a simple plot of $\Phi(s)$.}
\label{fig:phicdm}
\end{figure}

Hence, the parameters $L$
and $R$ are solutions of
the two following equations:
\begin{equation}
\label{cdm:constraints}
\frac{L^{3/2} }{12 \pi}+\frac{ \sqrt{L}}{\pi}=s\,\sqrt{b} \;\;\;
\textrm{and} \;\;\;  R=\frac{2 \pi}{\sqrt{L}}-\frac{ \pi \sqrt{L}}{2}
\end{equation}
These equations can be solved exactly.
In particular, we obtain the following expression
for $L=L(s)$:
\begin{eqnarray}
\label{L(s)}
L(s)&=&\left(-g_1(s)^{1/3} \, 2^{1/3}+ 
2^{5/3} \, g_1(s)^{-1/3}\right)^2\;\;\;\; \textrm{with} \;\; \;
g_1(s)=-3  \pi s\sqrt{b}+ \sqrt{16+9 b \pi^2 \, s^2}
\end{eqnarray}
The saddle point energy can then be computed 
(from equation  \eqref{cdmEffEner} and using
\eqref{cdm:SaddlePoint} for the calculation of the Lagrange multiplier $D$)
as a function of $L=L(s)$:
\begin{equation}
\label{EnMin}
E_s\left[\rho_c \right]=\frac{ L(s)^2}{32}- \ln\left(\frac{L(s)}{4}\right) +1
\end{equation}
Finally the distribution of the center of mass, 
in the large $N$ limit, is simply given by the steepest descent method
 $P(G_N=s \sqrt{N}) \propto
 \exp\left[-N^2 \, E_s\left[ \rho_c \right]\right]\;\;$:
\begin{equation}
\label{cdm:pdf}
P(G_N=\nu)\propto \exp\left[-N^2 \Phi\left( \frac{\nu}{\sqrt{N}}\right)\right]\;\;\;\;
\textrm{with} \;\;\; \Phi(s)=\frac{ L(s)^2}{32}- \ln\left(\frac{L(s)}{4}\right)
-\frac{1}{2}
\end{equation}
with $L=L(s)$ given in Eq. \eqref{L(s)}.
The additive constant has been chosen for convenience such that
the minimum of $\Phi$ is $0$. $\Phi$ is thus a positive function.
$\Phi(s)$ is plotted in Fig. \ref{fig:phicdm}.
As expected, the minimum of $\Phi(s)$
 is reached for $s=\mu$ , where $G_N=\mu\,\sqrt{N}
=\langle G_N \rangle$ -the average of the center of mass.
\\

\textbf{Validity of the regime where the density
has a support over $]0,L]$: $R\geq 0$, $s\leq \mu$}
\\

As we noticed above, the density $\rho_c$
 must be positive for every $0<x\leq L$, which is equivalent to
demanding that $R\geq 0$.
And from Eq. \eqref{cdm:constraints}, one can easily show
that the constraint $R \geq 0$ is equivalent to
$\mathbf{s \leq \mu}$. Thus the expression of $\Phi(s)$ given 
in Eq. \eqref{cdm:pdf}
is only valid on the left side of the mean of the center of mass:
$\nu \leq \mu \sqrt{N}$ (or $s \leq \mu$).
\\

\textbf{Limit $s\rightarrow\mu^-$ ($R\rightarrow 0^+$ ): 
Gaussian approximation of the pdf}

For $s \rightarrow \mu^-$, 
$\Phi$  can be expanded about its minimum:
\begin{equation}
\label{cdm=quadrexp}
\Phi(s)\approx \frac{(s-\mu)^2}{2 \sigma^2}\;\; {\rm where}
 \;\;
\mu=\frac{8}{3 \pi \sqrt{b}}\;\; {\rm and}\;\;
\sigma=\frac{1}{\pi} \sqrt{\frac{2}{b}}
\end{equation}
In this limit, the pdf of the center of mass can be
approximated by a Gaussian:
\begin{equation}
\label{cdmR+GaussApprox}
P(G_N=s \sqrt{N}) \propto e^{-  \frac{N^2\: (s-\mu)^2}{2 \sigma^2}} \;\;
\textrm{as} \;\; s \rightarrow \mu^-
\end{equation}
For large $N$, only the vicinity of $s=\mu$, 
where $\Phi$ is minimum, will contribute.
Therefore, the Gaussian approximation above
gives 
the mean value of the center of mass and its variance:
\begin{equation}
\label{cdmMin}
\langle G_N \rangle=\langle h \rangle\approx\mu \sqrt{N}
 \approx \frac{8}{3 \pi} \sqrt{\frac{N}{b}}
\end{equation}
 \begin{equation}
\label{cdmVar}
\hspace{-.2cm}\sqrt{\textrm{Var}(G_N)}=
\sqrt{\langle G_N^2\rangle-\langle G_N\rangle^2}
  \approx \frac{\sigma}{\sqrt{N}}
\approx 
\frac{1}{\pi}\sqrt{\frac{2}{N\, b}}
\end{equation}
This differs again strongly from the 
case of independent interfaces. 
For interfaces that are allowed to cross
(they are thus completely independent),
the average of the center of mass
 $\langle G_N \rangle=\langle h \rangle =m$ is of order one,
and its variance is given by
 $\sqrt{\textrm{Var}(G_N)}=\frac{\sigma_1}{\sqrt{N}}$, where
$m$ (resp. $\sigma_1$) is the mean (resp. variance)
of one single interface (see section \ref{1 int}). Both $m$ and $\sigma_1$
depend on $\alpha$ and $b$: they depend on the whole form of the potential
$V(h)=\frac{b^2 h^2}{2}+\frac{\alpha(\alpha-1)}{2 h^2}$.
But for nonintersecting interfaces, only the harmonic
part of the potential (with frequency $b$) has a non-negligible 
effect for large $N$ (the $\alpha$-dependence drops out, as 
we explained at the beginning of the section).
And the relative standard deviation 
$\frac{\sqrt{\textrm{Var}(G_N)}}{\langle G_N \rangle}$ is of 
 order $O\left(\frac{1}{\sqrt{N}} \right)$
for independent interfaces against
 $O\left(\frac{1}{N} \right)$ for nonintersecting interfaces.
The relative fluctuations are strongly reduced
by the fermionic repulsion.
\\

 \textbf{Limit $s\rightarrow 0^+ $ ($R \rightarrow +\infty$)  }
\\

For $s\rightarrow 0^+ $, the upper bound $L(s)$ of
the density support tends to zero like $s^2$:
 $L(s) \approx \pi^2\, s^2\;b 
+O(s^4) \;\; \textrm{as} \;\;  s\rightarrow 0^+$
and thus $\Phi$ tends to infinity :
\begin{equation}
\label{cdmPhiRinf}
\Phi(s)\approx -2 \ln s-\frac{1}{2}- \ln\left( \frac{\pi^2 b}{4}\right)
+O(s \ln s) \;\; \textrm{as} \;\;  s\rightarrow 0^+
\end{equation}
The probability density function thus tends to zero as a power law:
\begin{equation}
\label{cdmPdfRinf}
P(G=s \sqrt{N}) \propto s^{2 N^2} \;\; \textrm{as} \;\;  s\rightarrow 0^+
\end{equation}
\\

To summarize, for $s\le \mu=8/{\left(3\pi\sqrt{b}\right)}$,
 the large deviation function 
$\Phi(s)=\Phi^{-}(s)$
characterizing the form of the pdf of the center of mass $G_N$ to the left
of its mean value is given by Eqs. (\ref{cdm:pdf}) and (\ref{L(s)}),
and is plotted in Fig. \ref{fig:phicdm}. 
\\
\\

\subsubsection{Case $R<0$ ($s>\mu$)}

The previous regime (density with support over $]0,L]$) is only
valid for $R\geq 0$ or equivalently $s \leq \mu$. 
When $R<0$, the effective potential 
 is indeed minimal for
$x=x_0=\frac{R^2}{4 } >0$:
 the density $\rho_c$ is expected to have a finite support over $[L_1,L_2]$
with $L_1 >0$. $L_1$ and $L_2$ are 
 fixed by the constraints $\rho_c(L_1)=0=\rho_c(L_2)$.

In this subsection, we 
find an expression for
 $\rho_c$ when $R<0$ as a sum of elliptic integrals. 
We also derive the equations associated to the constraints
 $\rho_c(L_1)=0=\rho_c(L_2)$. But we could in general neither
 compute explicitely
the constraint $\int \sqrt{x} \rho_c(x) dx=s \, \sqrt{b}$
nor find a closed form for the energy (and $\Phi$),
except for the asymptotic regimes $s\rightarrow +\infty$ and
$s\rightarrow \mu^+$.
For $s\rightarrow \mu^+$, we show that
$\Phi(s)$ has  a very weak non-analyticity -an essential singularity-
at $s=\mu$: 
\begin{eqnarray}
\Phi^+(s) - \Phi^-(s)\approx
-\pi \, \sqrt{b}\,
(s-\mu)\,
 e^{-\frac{8}{\pi \, \sqrt{b} \, (s-\mu) }}\: e^{4 (\ln 2-1)}\: 
\;\;{\rm as }\;\; s\rightarrow \mu^+
\;\;{\rm where}\;\;
\Phi(s)=\left\{\begin{array}{l}
\Phi^-(s)\;\; {\rm for} \;\; s<\mu \\
\Phi^+(s)\;\; {\rm for} \;\; s>\mu
\end{array} \right.
\end{eqnarray}
\\

The (normalized) solution $\rho_c$, 
with support over $[L_1,L_2]$,
of the integral equation \eqref{cdm:hilbert}
is again given by Tricomi's theorem.
We get
\begin{eqnarray}
\label{cdmR-Dens}
 \rho_c(x)=\frac{1}{\pi^2 \sqrt{x-L_1} \sqrt{L_2-x}}
\Big[\pi+\frac{ \pi}{4}(L_1+L_2-2 x) \:
 +\frac{R\, \sqrt{L_2-L_1}}{4}\:
J\left(\frac{L_1}{L_2-L_1},\frac{x-L_1}{L_2-L_1}\right) 
 \Big] 
\end{eqnarray}
 with
\begin{eqnarray}
\label{cdmR-J}
J\left(\xi,y\right)&=&
\mathcal{P}\int_0^1 dt \frac{\sqrt{t} \, \sqrt{1-t}}{(t-y)\,
 \sqrt{t+\xi}}\nonumber \\
&=&-2 \sqrt{1+\xi} \: E\left(\frac{1}{1+\xi} \right)
 + \frac{2 \xi \sqrt{1+\xi}}{\xi+y} \: K\left(\frac{1}{1+\xi} \right)
-\frac{2 \xi (1-y)}{(\xi+y) \sqrt{1+\xi}} 
\: \: \Pi \left(\frac{\xi+y}{y (1+\xi)},\frac{1}{1+\xi} \right)
\end{eqnarray}
where $K$ and $E$ are the complete elliptic integrals
of the first and second kind respectively ; and  $\Pi$
is the incomplete elliptic integral of the
third kind:
\begin{equation}
\label{elliptic1}
E(k)=\int_0^1 \sqrt{\frac{1-k t^2}{1-t^2}} \, dt\;\;\;\textrm{and}  \;\;\;
K(k)=\int_0^1 \sqrt{\frac{1}{(1-k t^2)\, (1-t^2)}} \, dt
\end{equation}
\begin{equation}
\label{elliptic2}
\Pi(n,m)=\mathcal{P}
\int_0^1 \frac{1}{(1-n t^2)\,\sqrt{1-m t^2} \sqrt{1-t^2}} \, dt
\end{equation}

\begin{figure}[htp]
\centering
\begin{pspicture}(0,0)(6,6)
\includegraphics[width=7.8cm]{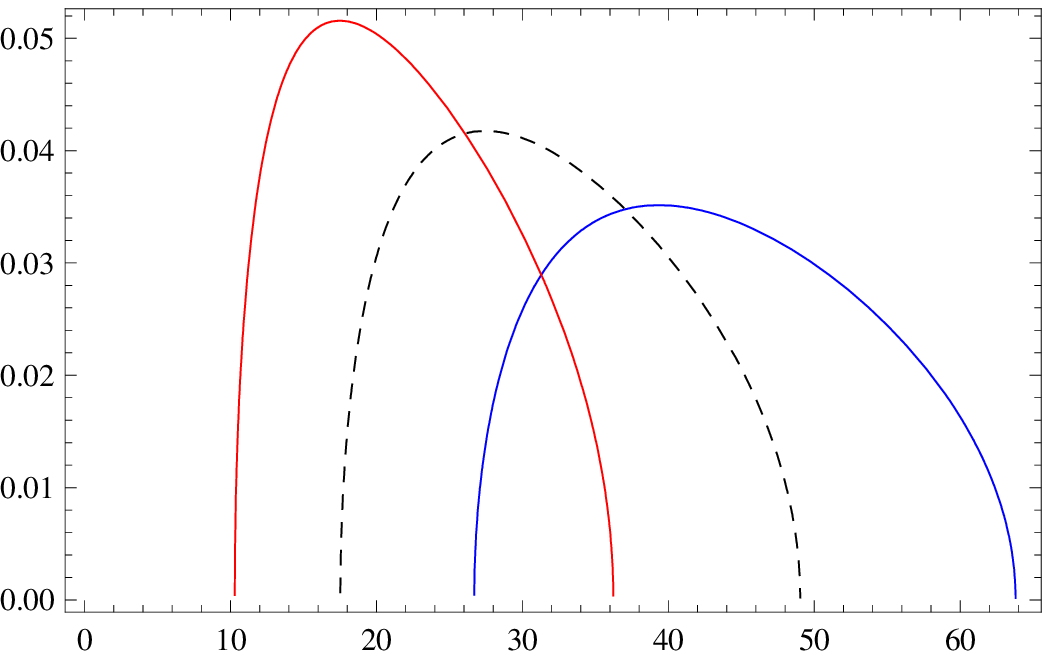}
\rput{0}(-1.2,3.1){\blue $R=-13$}
\rput{0}(-4.3,0.1){ \blue $L_1$}
\rput{0}(-.2,0.1){ \blue $L_2$}
\rput{0}(-3.4,4.1){ $R=-11$}
\rput{0}(-5.1,.6){ $L_1$}
\rput{0}(-1.7,.6){ $L_2$}
\rput{0}(-5.7,4){\red $R=-9$}
\rput{0}(-6.3,.6){\red $L_1$}
\rput{0}(-2.9,.6){\red $L_2$}
\rput{0}(-6.8,5.2){$\rho_c(x)$}
\rput{0}(-1.2,-0.2){$x$}
\end{pspicture}
\caption{
Density of states $\rho_c(x)$ (density of charges)
of the Coulomb gas associated to the computation of the pdf 
$P(G_N=s \sqrt{N})$ of the center of mass, 
in the case $s> \mu=\frac{8}{3 \pi \sqrt{b}}$ ($R<0$),
plotted for different values of the Lagrange multiplier $R$, or equivalently
different values of $s$ (and for $b=1$).
 The effective potential seen by the charges is minimal for 
$x=x_0=\frac{R^2}{4}>0$,
thus the density has a finite
support over $[L_1,L_2]$ and is maximal around $x=x_0$.
\\
 When $s>\mu$ and $s$ increases (i.e. the center
of mass is larger than its mean and increases),
$L_2>4$, $L_1>0$ and $L_2$ and $L_1$ increase also: the  charges
  form a bubble
that gets further from the origin when $R$ decreases
(or $s$ increases).}
\label{fig:dens R<0}
\end{figure}

We want to compute the pdf $P(G_N=s \,\sqrt{N})$. The basic variable
is thus $s$. There are now four unknown parameters: $R$ and $D$ are two
Lagrange
multipliers and $L_1$ and $L_2$  are the  bounds of the density support.
These parameters will be determined  by enforcing the four constraints
 $\int_0^{\infty} \rho(x)dx =1$,  $ \int_0^{\infty} \sqrt{x}
 \rho(x) dx = s\,\sqrt{b}$, $\rho_c(L_1)=0$ and $\rho_c(L_2)=0$.

For a given $R$ (Lagrange multiplier),
the parameters $L_1$ and $L_2$
are fixed by the constraints
$\rho_c(L_1)=0=\rho_c(L_2)$:
\begin{eqnarray}
\label{cdmR-Constraints}
\sqrt{L_2}&=&-\frac{R\: K(k)}{ \pi} \;\; \;\; \;\; 
\textrm{where} \;\; k=\frac{L_2-L_1}{L_2}=1-\frac{L_1}{L_2}\nonumber\\
\textrm{and}\;\;\;\;
\frac{2  \pi^2}{R^2}&=&-K(k)
 \left( E(k)+\left(\frac{k}{2}-1\right)\, K(k) \right)
\end{eqnarray}
We have already taken account of the constraint $\int  \rho_c(x) dx=1$
(normalization) by setting the constant $C_0$ that appears in Tricomi's
theorem (equation \eqref{cdmTricomi2})
 to $C_0=\pi \int \rho_c(x) dx=\pi$.

The  last constraint $\int \sqrt{x} \rho_c(x) dx=s\,\sqrt{b}$ gives
$R$ as a function of $s$. But the integral is in general
difficult to calculate.
And finally $D$ is in principle  given by the saddle point equation
(see Eq. \eqref{cdm:SaddlePoint})
at a special value of $x$, for example $x=L_1$. But
it is again difficult to compute in general.

Therefore we couldn't compute exactly the saddle point energy.
But, thanks to the above formulas, we could
plot the density for different values of $k$
(or equivalently, different values of $R$
or of $s$).
In this phase ($R<0$), as figure \ref{fig:dens R<0}
shows, the Coulomb charges accumulate 
in a band near the minimum
of the effective potential. They form a bubble
that gets further from the origin when $R$ decreases
(or $s$ increases). In this case, the interfaces are not
bound to the substrate.

We could also derive the asymptotics of $\Phi(s)$ in this regime:
$s \rightarrow +\infty$ and $s\rightarrow \mu^+$.
\\

\textbf{Right tail of the pdf: limit  $s \rightarrow + \infty$
($R \rightarrow -\infty)$}
\\

The limit  $R \rightarrow -\infty$
or equivalently $s \rightarrow + \infty$ corresponds
to $L_2 \rightarrow +\infty$ with 
$k=\frac{L_2-L_1}{L_2}\rightarrow 0^+$.
In this limit, we have
\begin{eqnarray}
\label{cdmR-infL1L2R}
R&\approx& \sqrt{2 } \left[ -\frac{8}{k}+4+\frac{21}{32} k+O(k^2)
\right]\;\; {\rm and} \;\;
\left\{\begin{array}{rcl}
L_2&\approx& \frac{32}{ k^2}-\frac{16}{k}-\frac{9}{4 }+O(k)\\
&&\\
L_1&\approx& \frac{32}{ k^2} -\frac{48}{k}+\frac{55}{4 }+O(k)
\end{array}\right.\;
\;\;{\rm as}\;\; k=1-\frac{L_1}{L_2}\rightarrow 0^+
\end{eqnarray}
And finally, for
$k=1-\frac{L_1}{L_2}\rightarrow 0^+ $ with 
$y=\frac{x-L_1}{L_2-L_1}$
fixed, $ 0<y<1$, we have:
\begin{equation}
\label{cdmR-infDens}
\rho_c(x) \approx
\frac{1}{4 \pi} \, \sqrt{y (1-y)} \:\; k \,+ O(k^2) 
\;\;{\rm with} \;\; y=\frac{x-L_1}{L_2-L_1}
\end{equation}
The constraint $\int \sqrt{x} \rho_c(x) dx=s \,\sqrt{b}$ gives
$s\approx \frac{4}{k}\sqrt{\frac{2}{b}}+O(1)$ as $k\rightarrow 0^+$,
and the minimal energy diverges:
\begin{equation}
\label{cdmR-infEnergy}
E_s[\rho_c]\approx \frac{32}{k^2}+O(\frac{1}{k})
 \;\; \textrm{as} \;\;k\rightarrow 0^+
\;\;\;{\rm thus} \;\; \Phi(s)\approx  s^2 b +O(s) \;\; \textrm{as} \;\;
s \rightarrow +\infty 
\end{equation}
which corresponds to a Gaussian tail:
\begin{equation}
\label{cdmR-infProb}
P(G_N=s \sqrt{N})\propto e^{-      b \, N^2 s^2     }  \;\; \textrm{as} \;\;
s \rightarrow +\infty 
\end{equation}

\subsubsection{Non-analyticity of the pdf:
limit  $s \rightarrow \mu^+$ ($R\rightarrow 0^-)$}

In this subsection, we analyse the limit $s\rightarrow\mu^+$,
which corresponds to $R\rightarrow 0^-$.

Let us define for convenience the following parameters:
\begin{eqnarray}
\label{def:xi}
\xi=\frac{L_1}{L_2-L_1} \;\;\;\;\;
(\xi\rightarrow 0 \;\;{\rm as}\;\; s\rightarrow\mu^+) \; \;\;\;\;\;
{\rm and}\; \;\;\;\;\;
X=-\frac{\ln\xi}{4}+\ln 2 \;\;\;\;\;
(X\rightarrow +\infty\;\; {\rm as}\;\; s\rightarrow\mu^+)
\end{eqnarray}
In the following, $\xi$ is chosen to be the small expansion parameter
(for $s\rightarrow\mu^+$). We will see that the expansion terms are of order
$O(X^{\eta} \, \xi^{\theta})=O\left( |\ln\xi|^{\eta}\,\xi^{\theta} \right)$
 with ${\theta}\geq 0$.
As $X^{\eta} \, \xi^{\theta} \gg X^{{\eta}'} \, \xi^{{\theta}'}$ 
($ |\ln\xi|^{\eta} \, \xi^{\theta} \gg  |\ln\xi|^{{\eta}'} \, \xi^{{\theta}'}$)
for $0\leq {\theta} < {\theta}'$ and for every ${\eta}$ and ${\eta}'$,
 we can make an expansion
in powers of $\xi$ of the form
$\sum_{{\theta}\geq 0} c_{\theta}(X) \, \xi^{\theta}$,
 where the exact value of the coefficients $c_{\theta}(X)$
 can be computed as functions 
of $X$ without expanding them. We thus keep 
 all the orders of the expansion in $X$ (expansion
in $\ln \xi$).

We will show that the saddle point energy (and thus the pdf of
the center of mass $P(G_N=s\sqrt{N})$) has a very weak
(infinite-order) non-analyticity at
$s=\mu$ (mean of the center of mass). More precisely, we will show that
the difference of the energy on the right and left side of
$\mu$ is of order $ O\left(\frac{\xi}{X}\right)
 \approx O\left(\frac{\xi}{|\ln \xi|}\right)
\approx O\left( |s-\mu| \, e^{-\frac{8}{\pi\sqrt{b}\,|s-\mu|}}\right)$:
 it is an essential singularity
(it is much smaller than any power of $| s-\mu|$).
\\

 \textbf{A singular limit for the saddle point density}

Using the  equations  \eqref{cdmR-Constraints} obtained by enforcing the
constraint $\rho_c(L_1)=0=\rho_c(L_2)$, we can expand the Lagrange multiplier
$R$ and the bounds $L_1$ and $L_2$ in terms of the small parameter
$\xi=\frac{L_1}{L_2-L_1}$, to first order in $\xi$:
\begin{eqnarray}
\label{cdmNonAn:RandL}
R\approx \frac{-\pi}{\sqrt{X(X-1)}} 
 +\xi \: \left( \frac{\pi \, (4 X^2+2 X-1)}{16\, \left[X(X-1)\right]^{3/2}}
\right) +O\left(\frac{\xi^2}{X}\right)\;\; {\rm and}\;\;
\left\{\begin{array}{rcl}
L_2&\approx&\left( \frac{4 X}{X-1}  \right)+\xi \:
\left( -\, \frac{(4X +1)}{2 (X-1)^2} \right)+O(\xi^2)\;\;\;\;\;\;\\
&&\\
L_1&\approx&\xi\: \left( \frac{4 X}{X-1} \right)+O(\xi^2)
\end{array}\right.
\end{eqnarray}
with  $\xi=\frac{L_1}{L_2-L_1}$ and $X=-\frac{\ln\xi}{4}+\ln 2$.
\vspace{.2cm}

For $s\rightarrow \mu^+$, we have
$\xi\rightarrow 0$ ($X\rightarrow +\infty$) and we recover
$R\rightarrow 0$ (with $R<0$), $L_2\rightarrow 4$ and $L_1\rightarrow 0$.
These are the same limits as on the left side of $\mu$:
for $s\rightarrow \mu^-$, we have $R\rightarrow 0$ and the density
has a support over $]0,L]$ with $L\rightarrow 4$.
\\

The saddle point density is given by Tricomi's theorem
in equation  \eqref{cdmR-Dens}. Using the constraint $\rho_c(L_1)=0$,
we get
\begin{eqnarray}
\label{cdmNonAn:rhoc}
\rho_c(x)=\rho(y)\equiv 
A\, \sqrt{\frac{y}{1-y}} +\frac{B}{\sqrt{y (1-y)}}
+\frac{C  \, J(\xi,y)}{\sqrt{y (1-y)}}
\;\;\;\;\;\;\;
{\rm with }\;\; y=\frac{x-L_1}{L_2-L_1} \;\; (0\leq y\leq 1) 
\end{eqnarray}
where $J(\xi,y)$
can be expressed as the principal value of an integral
(see equation \eqref{cdmR-J}):
\begin{equation}
\label{cdmNonAn:J}
J(\xi,y)=\mathcal{P}\int_0^1 dt \frac{\sqrt{t} \, \sqrt{1-t}}{(t-y)\,
 \sqrt{t+\xi}}
\end{equation}
and where the coefficients $A=-\frac{1}{2 \, \pi}$, 
$B=\frac{-R\, J(\xi,0)}{4 \pi^2 \sqrt{L_2-L_1}}$ and 
$C=\frac{R}{4 \pi^2 \sqrt{L_2-L_1}}$ can easily be expanded to first order in
$\xi$.
\\

For $y \in ]0,1[$ fixed and for $\xi\rightarrow 0$
($s\rightarrow\mu^+$), we have
\begin{eqnarray}
\label{cdmNonAn:Jexpand}
J(\xi,y)\approx -2+2 \sqrt{1-y}\, \argtanh(\sqrt{1-y})-
\frac{\xi\,\ln\xi}{2 y} +O(\xi)\;\;\;\; {\rm and} \;\;\;\;
 L_1\approx O(\xi)\;\;\;\;
{\rm as}\;\; 
\xi \rightarrow 0\;\;
\end{eqnarray}
Therefore,  to zeroth order in $\xi$, the density shape
(for $L_1 < x<L_2$) is the
same as for $s<\mu$, it diverges for small $x$:
\begin{eqnarray}
\label{cdmNonAn:rhocExpand}
\rho_c(x)\approx\frac{1}{2\pi}\sqrt{\frac{L_2-x}{x}}-\frac{1}{4 \pi X}\,
\sqrt{\frac{L_2}{x}}\, \argtanh\left(\sqrt{1-\frac{x}{L_2}}\right)+O(\xi)
\end{eqnarray}
But (for $s>\mu$), the density $\rho_c(x)$ has a finite support
over $[L_1,L_2]$ with $L_1>0$: it must vanish at $x=L_1$.
The constraint $\rho_c(L_1)=0$ seems to be violated
in Eq. \eqref{cdmNonAn:rhocExpand}, but it is not. 
As $L_1 \approx O(\xi)$, the part of the density associated to
small $x$ (close to $L_1$) - and where the density must approach zero-
does indeed not contribute to the zeroth order expansion of the density.
The weight of the 
small range of values of $x$ (around $L_1$) where the density
grows from zero to a very large value  just becomes negligible
when $s\rightarrow\mu^+$.

The limiting shape of the density for
$s\rightarrow \mu^+$ is thus singular.  
Therefore it is  better not to 
 expand $J(\xi,y)$ and the density for fixed $y$ (fixed $x$) and small $\xi$,
but to directly make an expansion of the energy,
that involves integrals such that $\int dy J(\xi,y) \sqrt{y+\xi}$
or $\int dy J(\xi,y) \ln y$.
Otherwise,  as the
 limits $y\rightarrow 0$ and  $\xi\rightarrow 0$ do not commute,
  the expansion of
$J(\xi,y)$ in terms of powers of $\xi$
will generate increasing negative  powers of $y$ that will
make  integrals like $\int dy J(\xi,y) \ln y$ diverge in zero.
\\

\textbf{Expansion of the constraint $\int dx \rho_c(x) \sqrt{x}=s\sqrt{b}$
for $s\rightarrow \mu^+$}

We must enforce the constraint
 $\int_{L_1}^{L_2} dx \rho_c(x) \sqrt{x}=s\sqrt{b}$
that replaces the delta function 
$\delta\left(\frac{\sqrt{\lambda_1}+...+\sqrt{\lambda_N}}{N \, \sqrt{b}}-
s \sqrt{N}\right)$ in the expresssion of the pdf of the center of mass
$P(G_N=s \sqrt{N})$:
\begin{eqnarray}
\label{cdmNonAn:constraintSqrt}
s \, \sqrt{b}&=&\int_{L_1}^{L_2} dx \rho_c(x) \sqrt{x}=
(L_2-L_1)^{3/2} \, \int_0^1 dy \rho(y)\, \sqrt{y+\xi}
\end{eqnarray} 
From the expression of $\rho_c$ given in Eq. \eqref{cdmNonAn:rhoc}, we 
see that we need to expand for small $\xi$ a double 
improper integral:
\begin{equation}
\label{cdmNonAn:defI}
I(\xi)=\int_0^1 dy \frac{ \sqrt{y+\xi} \, J(\xi,y)}{\sqrt{y (1-y)}}
=\int_0^1 dy \frac{\sqrt{y+\xi}}{\sqrt{y (1-y)}} \,
 \mathcal{P}\int_0^1 dt\,
\frac{\sqrt{t (1-t)}}{\sqrt{t+\xi}} \, \frac{1}{t-y}
\end{equation}
As $I(\xi)$ is a double improper integral
(with principal value), it is not easy to compute it or even expand it
directly (for small $\xi$). Let us first make a simple transformation
in order to get rid of the principal value:
\begin{eqnarray}
\label{cdmNonAn:Icalc}
I(\xi)
=I(\xi=0)+\xi \,f_0(\xi)  
\;\; {\rm with} \;\; 
f_0(\xi)= \int_0^1 dy \, \int_0^1 dt\,
\frac{\sqrt{ (1-t)}}{\sqrt{y (1-y)}} \, \frac{1}{\sqrt{t+\xi} \left[
\sqrt{t(y+\xi)}+\sqrt{y(t+\xi)} \right]}
\end{eqnarray}
where $I(\xi=0)=-2$ (it can be easily computed exactly)
and  where $f_0$ is a definite double integral, easier to expand.
However, as we already noticed,
the limit $\xi\rightarrow 0$ and the integration do not commute:
the expansion can not be done inside the integral.
Hence, the method of expansion must be a bit more subtle.
Our method (see appendix-B for details) consists in splitting the
 initial integral $f_0(\xi)$
in a sum of integrals (some of them are easier to compute,
the other ones are shown to be negligible).

Finally (see appendix-B)
 we get the expansion of $I(\xi)$ to first order in $\xi$
(but to all orders in $X$, or $\ln \xi$):
\begin{eqnarray}
I(\xi)
&\approx&-2 +\xi \:   \Big[
8 X^2-4 X-1
\Big]
+O(\xi^2\, X^2)
\;\;\;\;\;\;\;\;\;\;\;\; {\rm as }\;\;
\xi\rightarrow 0   \label{cdmNonAn:IExpand}
\end{eqnarray}
Hence the constraint $\int dx \rho_c(x) \sqrt{x}=s\sqrt{b}$
is given by
\begin{eqnarray}
\label{yracb}
s \, \sqrt{b}&\approx&\frac{2 \, (4 X-3)\, \sqrt{X}}{3 \pi \, (X-1)^{3/2}}
+\xi \: \left( \frac{-16 X^3+12 X^2-2 X+1}{8 \pi\, 
(X-1)^{5/2}\, \sqrt{X}}\right)+O(\xi^2\, X) \;\;\;\;\;\; {\rm as }\;\;
\xi\rightarrow 0
\end{eqnarray}
In particular, as expected,
when $\xi\rightarrow 0^+$ ($X\rightarrow \infty$),
 $s\,\sqrt{b}$ tends to
the mean value $\mu \sqrt{b}=\frac{8}{3\pi}$.

Moreover, the formula above (Eq. \eqref{yracb}) can be inverted to
express $X$ and $\xi$ as functions of $(s-\mu)$.
As $\mu\,\sqrt{b}=\frac{8}{3\pi}$, we get:
\begin{equation}
\label{eq:xi}
X\approx \frac{2}{\pi \, (s-\mu)\,\sqrt{b}}+1+O\left( (s-\mu)\right)
\;\;\;{\rm and}\;\;\;
\xi\approx e^{\frac{-8 }{\pi \, (s-\mu)\,\sqrt{b}}}
\;\; e^{4 \,(\ln 2-1)} \; \left(1+O\left( (s-\mu)\right) \right)
\;\; {\rm as} \;\; s\rightarrow\mu^+
 \end{equation}
\\

\textbf{Energy $E_s[\rho_c]$
and scaling function $\Phi(s)=E_s[\rho_c]-\frac{3}{2}$}

From equation \eqref{cdmEffEner},
we can compute the saddle point energy: 
\begin{equation}
E_s[\rho_c]=\frac{1}{2}\int_{L_1}^{L_2}dx \, \rho_c(x)  \,x 
-\frac{R}{2} \, s\, \sqrt{b}-\frac{D}{2}
\end{equation}
where the Lagrange multiplier $D$ can be calculated by replacing
$x$ by $L_1$ in the saddle point equation for the density
(equation \eqref{cdm:SaddlePoint})
and where $ \int_{L_1}^{L_2}dx \, \rho_c(x)  \,x$ 
is not very difficult to expand for  $\xi\rightarrow 0$.
Finally we get the expression of the  energy
for $s\rightarrow \mu^+$
($\xi\rightarrow 0$), to first order in $\xi$
and all orders in  $X=\ln 2-\frac{\ln\xi}{4}$:
\begin{equation}
\label{eq:energyOrdre2}
E_s[\rho_c]\approx \ln\left(\frac{X-1}{X} \right)+
\left(
\frac{3 X^2-4 X+2}{2 \, (X-1)^2}\right)+\xi\: 
\left(\frac{-16 X^3+16 X^2-6 X+1}{8 X\, (X-1)^3}\right)+O(\xi^2)
\end{equation}
Using equation \eqref{eq:xi} giving the expression of $\xi$
 as a funtion of $(s-\mu)$ in the limit
$s\rightarrow \mu^+$, we will thus derive the behavior of
the pdf of the center of mass $P(G_N=s\sqrt{N})
 \propto e^{-N^2 E_s[\rho_c]}$ (for large $N$)
for $s\rightarrow \mu^+$.

In order to show that the pdf of the center of mass has
 a non-analyticity at $s=\mu$, we must compare the expansion of
the saddle point energy on the left side and the right side of the mean.
 \\

\textbf{Zeroth order in $\xi$: $\Phi(s)$ seems to be a smooth function}

 Let us first consider the zeroth order in the expansion
in terms of powers of $\xi$ (on the right side of $\mu$).
To zeroth order in $\xi$, the constraint 
 $\int dx \rho_c(x) \sqrt{x}=s\sqrt{b}$  given in
Eq. \eqref{yracb} reduces to
\begin{eqnarray}
\label{eq:L2}
s \, \sqrt{b}&\approx&\frac{2 \, (4 X-3)\, \sqrt{X}}{3 \pi \, (X-1)^{3/2}}
+O(\xi)
\approx \frac{L_2^{3/2}}{12 \pi}+\frac{ L_2^{1/2}}{\pi}+O(\xi)
\end{eqnarray}
Therefore, to all orders in $\ln \xi$
(or $X$ -but to zeroth order in
$\xi$), we recover  the same equation as
Eq. \eqref{cdm:constraints}, i.e. the same equation as
on the left side of the mean, giving $L$ ($L_2$) as a function of $s$!
The Lagrange multiplier $R$ is also given, to zeroth order in
$\xi$ by the same function
of $L_2$ ($L$) as on the left side of the mean
(see Eq. \eqref{cdm:constraints}):
\begin{eqnarray}
\label{cdmNonAn:Rord0}
\frac{2\pi}{\sqrt{L_2}}-\frac{\pi\sqrt{L_2}}{2}\approx
\frac{-\pi}{\sqrt{X(X-1)}} 
+O(\xi)\approx R +O(\xi)
\end{eqnarray}

Finally, the energy, to zeroth order in $\xi$
(but to all orders in $X$ or $\ln \xi$)
is given by the same expression as the energy on the
 left side of the mean:
\begin{eqnarray}
E_s[\rho_c]^+&\approx&
 \ln\left(\frac{X-1}{X} \right)+
\left(
\frac{3 X^2-4 X+2}{2 \, (X-1)^2}\right)+O\left(\frac{\xi}{X}\right)\nonumber \\
&\approx&\frac{L_2^2}{32}-2 \ln\left(\frac{\sqrt{L_2}}{2}\right)+1+
O\left(\frac{\xi}{X}\right)\nonumber \\
&\approx& E_s[\rho_c]^- +
O\left(\frac{\xi}{X}\right) \label{cdmNonAn:energyDiff}
\end{eqnarray}
where $L_2=L_2(s)$ is given by equation  \eqref{eq:L2},
the same equation for $s\rightarrow\mu^+$ to zeroth order
in $\xi$ as for $s\rightarrow\mu^-$.
As 
$\xi\approx e^{\frac{-8 }{\pi \, (s-\mu)\,\sqrt{b}}}
\;\; e^{4 \,(\ln 2-1)}$
when $ s\rightarrow\mu^+$ (see equation \eqref{eq:xi}), we get:
\begin{eqnarray}
\Phi^+(s)-\Phi^-(s)=E_s[\rho_c]^+ -E_s[\rho_c]^- \approx
O\left(\frac{\xi}{X}\right)\approx O\left(|s-\mu| \:
 e^{\frac{-8 }{\pi \,  |s-\mu|\,\sqrt{b}}}
\right)  \label{cdmNonAn:PhiDiff}
\end{eqnarray}

All the terms of the
 expansion of the energy (and thus $\Phi(s)$ and the pdf of the center of mass)
in powers of $|s-\mu|$ (or $\frac{1}{\ln\xi}$ or $\frac{1}{X}$)
are thus the same on the left and right side of the mean:
$\Phi(s)$ is a smooth function, it is infinitely differentiable
even at $s=\mu$ -in particular the quadratic approximation
of $\Phi(s)$ in Eq. \eqref{cdm=quadrexp}
 is valid on both left and right
side of its minimum ($s=\mu$).
However, we will show that the expansion to first order in $\xi$
(by keeping all the powers of $X$) gives 
  a very weak non-analyticity of the energy (and thus $\Phi(s)$).
\\

\textbf{First order in $\xi$: non-analyticity of $\Phi(s)$}

Using equation \eqref{eq:energyOrdre2} and the remarks we made
about the zeroth order expansion in $\xi$, 
we get the difference between the expansion of
the energy on the right and left side of $\mu$:
\begin{eqnarray}
 \label{cdmNonAn:energyDifford1}
E_s[\rho_c]^+ -E_s[\rho_c]^-
&\approx& \xi\: 
\left(\frac{-16 X^3+16 X^2-6 X+1}{8 X\, (X-1)^3}\right)+O(\xi^2)
\end{eqnarray}
Using the expression of $\xi$ and 
$X= \ln 2-\frac{\ln \xi}{4}$
as function of $s$
for $s\rightarrow\mu^+$ given in Eq. \eqref{eq:xi},
 we finally get
\begin{equation}
 \label{cdmNonAn:PhiNonAnalyticity}
\Phi^+(s)-\Phi^-(s)=E_s[\rho_c]^+ -E_s[\rho_c]^-
\approx
-\pi \, \sqrt{b}\,
(s-\mu)\,
 e^{-\frac{8}{\pi \, \sqrt{b} \, (s-\mu) }}\: e^{4 (\ln 2-1)}\: 
\;\;{\rm as }\;\; s\rightarrow \mu^+
\end{equation}
This is an essential singularity.
We have  shown that
the pdf of the center of mass
$P(G_N=s \,\sqrt{N})\propto e^{-N^2 E_s[\rho_c]}$
 has a very weak 
non-analyticity at $s=\mu$: the energy (or equivalently $\Phi(s)$)
has an infinite-order non-analyticity,
of order
$O\left( |s-\mu|\,
 e^{-\frac{8}{\pi \, \sqrt{b} \, |s-\mu| }}\right)$.

\section{Conclusion}
\label{sec:ccl}

In summary, we have studied a simple model of $N$ nonintersecting
fluctuating interfaces at thermal equilibrium and in presence
of a wall that induces an external confining potential of the form
$V(h)=\frac{b^2 h^2}{2}+\frac{\alpha(\alpha-1)}{2 h^2}$. Our study extablishes
a deep connection between the statistics of heights of the interfaces
in the limit of a large system ($L\rightarrow\infty$)
 and the eigenvalues of the Wishart random matrix,
thus providing a nice and simple physical realization of the Wishart
ensemble.
More precisely, we have  proved that the joint probability distribution
of the interface heights $h_i$ in the limit of a large system can be mapped to
 the distribution of the eigenvalues $\lambda_i$
of a Wishart matrix under the change of variables 
$b\, h_i^2=\lambda_i$, with arbitrary parameter value $M-N$ of the
Wishart ensemble that is fixed by the parameter $\alpha$ of the
inverse square external potential. 

We have also shown how to exploit the relation
between interfaces and eigenvalues of the Wishart matrix
to derive asymptotically exact results for the height statistics 
in the interface model.
In particular, we have seen that the nonintersecting constraint,
the only interaction between interfaces in our model,
drastically changes the behavior of interfaces:
they become strongly correlated. Despite the presence of
strong correlations that make the problem difficult
to analyse, we were able to compute a number of
asymptotic (large $N$) results exactly. These include the
computation of the average density of states, the distribution
of maximal and minimal heights and the distribution of the
center of mass of the interfaces. In the last case, we have shown
that the distribution has an extraordinarily weak singularity 
near its peak (an essential singularity)
and this non-analytical behavior was shown to be a direct
consequence of a phase transition in the associated Coulomb gas problem.

Finally, we expect that the appearence of the Wishart random matrix
in a physically realizable example as shown in this paper will be
useful in other contexts. In addition, the Coulomb gas technique 
used here seems to be a very nice way to derive exact asymptotic results
in this class of interacting many body systems where exact analytical
results are hard to come by. 
It would be interesting to use the analogy with a Coulomb gas
in other physical problems related to Wishart matrices,
for example to compute the distribution of entropy of
a bipartite quantum system (see \cite{majumdar-quant,facchi}).

\vskip 0.2cm

\noindent {\bf Acknowledgements:} It is a pleasure to thank A. Comtet for many useful discussions.

\appendix
\section{Computation of the moments of the minimal height}

 \eqref{eq:minPDF} gives an exact expression for
the pdf of the minimal height (lowest interface):
\begin{equation}
\label{eq:minPDF1}
P\left(h_{\rm min}= t, N \right)
=2 \,b^2 \,  t^3 \,  e^{-b N t^2} \:
\mathcal{L}_{N-1}^{(2)}(-b\, t^2)
=b^2 \,  t^3 \,  e^{-b N t^2} \: N(N+1) \,_1F_1(1-N,3,-b t^2)
\end{equation}
(see \eqref{hyper1} for the relation
between Laguerre polynomials and hypergeometric functions)

Therefore we can compute explicitely the moments of the minimal height:
\begin{eqnarray}
\label{eq:minPDF2}
\langle h_{\rm min}^k \rangle&=& \int_0^{\infty}
 dt \: t^k \, P\left(h_{\rm min}= t, N \right)
= b^2\,N(N+1)\, \int_0^{\infty} dt \: t^{k+3}\:  e^{-b N t^2} \:
  \,_1F_1(1-N,3,-b t^2)\\
&=& b^2\,N(N+1)\, \frac{b^{-k/2-2}}{2}
\, \int_0^{\infty} du \: u^{k/2+1}\:  e^{- N u} \:
  \,_1F_1(1-N,3,-u)
\end{eqnarray}
with $b t^2=u$.
The integral above can be computed:
$\int_0^{\infty}du \: u^{d-1} \, e^{-cu}\, 
\,_1F_1(a,b,-u)=c^{-d} \, \Gamma(d)\, \,_2F_1(a,d;b;-1/c)$

Therefore
\begin{equation}
\langle h_{\rm min}^k \rangle=
\frac{\Gamma(k/2+2)}{2\, b^{k/2}}\,\frac{(N+1)}{N^{k/2+1}}\:
\,_2F_1(1-N,k/2+2;3;-1/N)
\end{equation}

For example, for $k=1$, we find:
\begin{equation}
\langle h_{\rm min} \rangle=
\frac{\Gamma(5/2)}{2\, b^{1/2}}\,\frac{(N+1)}{N^{3/2}}\:
\,_2F_1(1-N,5/2;3;-1/N)
\end{equation}
For large $N$, as $ _2F_1(1-N,5/2;3;-1/N)
=\sum_n \frac{(5/2)(7/2)...(3/2+n)}{\,(3)(4)...(n+2)\: n!}\, 
\frac{(1-N)(2-N)...(n-N)}{(-N)^n}$, we find:
\begin{eqnarray*}
\lim_{N\rightarrow \infty}\,_2F_1(1-N,5/2;3;-1/N)&=&
\sum_n \frac{(5/2)(7/2)...(3/2+n)}{(3)(4)...(n+2)\: n!}\\
&=&_1F_1(5/2;3;1)=\frac{4 \sqrt{e}}{3} I_0(1/2)
\end{eqnarray*}
Hence, for large $N$:
\begin{equation}
\langle h_{\rm min} \rangle \approx \frac{c_1}{ \sqrt{b\,N}}
\end{equation}
with
\begin{equation}
c_1=\frac{\Gamma(5/2)}{2} \frac{4 \sqrt{e}}{3} \: I_0(1/2)=
 \sqrt{\frac{\pi e}{4}} \: I_0(1/2)\approx 1.5538
\end{equation}

\section{Non-analyticity of the pdf of
the center of mass: expansion of $I(\xi)$}

Let us expand  for $\xi\rightarrow 0$ the integral $I(\xi)$
given in Eq. \eqref{cdmNonAn:defI}:
\begin{equation}
\label{B:defI}
I(\xi)=\int_0^1 dy \frac{ \sqrt{y+\xi} \, J(\xi,y)}{\sqrt{y (1-y)}}
\end{equation}
As $I(\xi)$ is a double improper integral
(with principal value), it is not easy to compute it or even expand it
directly (for small $\xi$). Let us first make a simple transformation
in order to get rid of the principal value:
\begin{eqnarray}
I(\xi)&=&\int_0^1 dy \frac{\sqrt{y+\xi}}{\sqrt{y (1-y)}} \,
 \mathcal{P}\int_0^1 dt\,
\frac{\sqrt{t (1-t)}}{\sqrt{t+\xi}} \, \frac{1}{t-y}\nonumber\\
&=&I(\xi=0)+\int_0^1 dy \, \mathcal{P}\int_0^1 dt\,
\frac{\sqrt{t (1-t)}}{\sqrt{y (1-y)}} \, \frac{1}{t-y}\,
\left( \sqrt{\frac{y+\xi}{t+\xi}} -\sqrt{\frac{y}{t}}\right)\nonumber\\
&=&-2+\xi \int_0^1 dy \, \int_0^1 dt\,
\frac{\sqrt{ (1-t)}}{\sqrt{y (1-y)}} \, \frac{1}{\sqrt{t+\xi} \left[
\sqrt{t(y+\xi)}+\sqrt{y(t+\xi)} \right]}\nonumber\\
&\equiv&-2+\xi \,f_0(\xi) \label{B:Icalc}
\end{eqnarray}
The value of $I(\xi=0)$ can indeed be computed exactly:
\begin{eqnarray}
I(\xi=0)=\int_0^1 dy \frac{1}{\sqrt{ 1-y}} \,
 \mathcal{P}\int_0^1 dt\,
\frac{\sqrt{1-t}}{ t-y}
=\int_0^1 \frac{dy}{\sqrt{1-y}} \left[ -2
+2 \sqrt{1-y}\: \argtanh\left(\sqrt{1-y}\right)   \right] 
=\,-2  \label{B:I(xi=0)}
\end{eqnarray}
 $f_0$ is a definite double integral, easier to expand:
\begin{eqnarray}
\label{B:f0}
f_0(\xi)= \int_0^1 dy \, \int_0^1 dt\,
\frac{\sqrt{ (1-t)}}{\sqrt{y (1-y)}} \, \frac{1}{\sqrt{t+\xi} \left[
\sqrt{t(y+\xi)}+\sqrt{y(t+\xi)} \right]}
\end{eqnarray}
We thus need to expand a definite double integral $f_0(\xi)$
(we have got rid of the principal value). 

However, as we already noticed,
the limit $\xi\rightarrow 0$ and the integration do not commute:
the expansion can not be done inside the integral.

Let us thus consider separately the integration
over $]0,\xi]$ and $[\xi,1[$ (for the variable $y$):
\begin{equation}
\label{B:Isplit}
 f_0(\xi)=f_1(\xi)+f_2(\xi)
\end{equation}
where $f_1$ and $f_2$ are definite double integrals
(no principal value):
\begin{eqnarray}
f_1(\xi)&=& \int_0^{\xi} dy \, \int_0^1 dt\,
\frac{\sqrt{ (1-t)}}{\sqrt{y (1-y)}} \, \frac{1}{\sqrt{t+\xi} \left[
\sqrt{t(y+\xi)}+\sqrt{y(t+\xi)} \right]}\label{B:I1def}\\
f_2(\xi)&=& \int_{\xi}^1 dy \, \int_0^1 dt\,
\frac{\sqrt{ (1-t)}}{\sqrt{y (1-y)}} \, \frac{1}{\sqrt{t+\xi} \left[
\sqrt{t(y+\xi)}+\sqrt{y(t+\xi)} \right]}\label{B:I2def}
\end{eqnarray}

The method of expansion will be the following.
We will write $f_1$ (resp. $f_2$) as a sum of two integrals:
the first will be chosen of the same order of $f_1$ (resp. $f_2$)
for small $\xi$,
but easier to compute 
(by separation of the variables $t$ and $y$ for example);
the second will be  much smaller than the first and than $f_1$ (resp. $f_2$).
Then, following the same scheme,
each of these two integrals can again be split into two 
pieces, until we get the full expansion to first order in $\xi$
(the last integral will be shown to be much smaller than the other 
and will be neglected).

 Let us first make the change of variable $y=\xi\, u$ in $f_1(\xi)$:
\begin{eqnarray}
\label{B:I1changeVar}
f_1(\xi)=\int_0^1 du \int_0^1 dt \frac{\sqrt{1-t}}{\sqrt{u} \, \sqrt{1-\xi u}\,
 \sqrt{t+\xi} }\, \frac{1}{\left( \sqrt{t(1+u)}+\sqrt{u (t+\xi)} \right)}
\end{eqnarray}
Then we have
\begin{eqnarray}
\label{B:I1split}
f_1(\xi)=f_1^a(\xi)+f_1^b(\xi)
\end{eqnarray}
with $f_1^a$ a product of two integrals (separation of variables)
\begin{eqnarray}
f_1^a(\xi)&=&\left( \int_0^1 \frac{du}{\sqrt{u}\,(\sqrt{u}+\sqrt{1+u}) \,
 \sqrt{1-\xi u} }\, \right) \, \left( \int_0^1 dt
\frac{\sqrt{1-t}}{ \sqrt{t} \,  \sqrt{t+\xi}}\right) \nonumber\\
&\approx&\left(  \int_0^1 \frac{du}{\sqrt{u}\,(\sqrt{u}+\sqrt{1+u}) \,
 }\,+O(\xi) \right) \, \left( \int_0^1 dt
\frac{\sqrt{1-t}}{ \sqrt{t} \,  \sqrt{t+\xi}}\right) \nonumber\\
&\approx&\left(-1+\sqrt{2}+\argsinh 1 +O(\xi) \right) \, \left(
-\ln \xi +4 \ln 2 -2 +O\left(\xi \,\ln\xi \right) \right) \nonumber\\
&\approx& \ln \xi \, \left(  1-\sqrt{2}-\argsinh 1
\right)+(4 \ln 2-2 )(-1+\sqrt{2}+\argsinh 1)+O\left(\xi \, \ln \xi \right)
\label{B:I1a}
\end{eqnarray}
and $f_1^b=f_1-f_1^a$ (expected to be much smaller
than $f_1$) is given by:
\begin{eqnarray}
\hspace{-1cm}
f_1^b(\xi)&=&\int_0^1 du \int_0^1 dt \frac{\sqrt{1-t}}{\sqrt{u }\, 
\sqrt{1-\xi u}\,
 \sqrt{t+\xi} }\,
\left[\frac{1}{ \left( \sqrt{t(1+u)}+\sqrt{u (t+\xi)} \right)}-
\,\frac{1}{\sqrt{t} \, (\sqrt{u}+\sqrt{1+u})}
\right]\nonumber \\
&=&-\xi\, \int_0^1 \frac{du}{\sqrt{1-\xi \, u}\, (\sqrt{u}+\sqrt{1+u})}
\int_0^1 dt \frac{\sqrt{1-t}}{\sqrt{t (t+\xi)} \, (\sqrt{t}+\sqrt{t+\xi})
\,\left( \sqrt{t (1+u)}+\sqrt{u (t+\xi)}\right)}\nonumber\\
&=&f_1^c(\xi)+f_1^d(\xi) \label{B:I1bsplit}
\end{eqnarray}
with
\begin{eqnarray}
\hspace{-1cm}
f_1^c(\xi)&=&
-\xi\, \int_0^1 \frac{du}{\sqrt{1-\xi \, u}\, (\sqrt{u}+\sqrt{1+u})}
\int_0^1 dt \frac{\sqrt{1-t}}{\sqrt{t (t+\xi)} \, (\sqrt{t}+\sqrt{t+\xi})
\,\left( \sqrt{ u}+\sqrt{1+u} \right) \,\sqrt{t+\xi}}\nonumber\\
&=&-\left(\int_0^1 \frac{du}{\sqrt{1-\xi \, u}\, (\sqrt{u}+\sqrt{1+u})^2}
\right)\,
\left(\int_0^{1/\xi}dz \frac{\sqrt{1-\xi \, z}}{(1+z)\sqrt{z} 
(\sqrt{z}+\sqrt{1+z})}  \right)\nonumber \\
&\approx& -\left(\int_0^1 \frac{du}{\sqrt{ u}\, (\sqrt{u}+\sqrt{1+u})^2}
+O(\xi)
\right)\,
\left(\int_0^{\infty }  \frac{dz}{(1+z)\sqrt{z} 
(\sqrt{z}+\sqrt{1+z})} +O(\xi \,\ln\xi) \right) \nonumber\\
&\approx& \left( \frac{3}{\sqrt{2}}-2-\frac{\argsinh 1}{2}+O(\xi) \right)
\, \left( 2 \ln 2+O(\xi\,\ln\xi) \right)\nonumber\\
&\approx& \ln 2 \left( 3\sqrt{2}-4-\argsinh 1 \right) +O(\xi\, \ln\xi) 
\label{B:I1c}
\end{eqnarray}
where we have made the change of variables  $t=\xi \,z$;
and
\begin{eqnarray}
\hspace{-3cm}
f_1^d(\xi)&=&
-\xi^2 \int_0^1 \frac{du\, \sqrt{1+u}}{
\sqrt{1-\xi \, u}\, (\sqrt{u}+\sqrt{1+u})^2}\,
\int_0^1  \frac{dt\; \sqrt{1-t}}{\sqrt{t} (t+\xi) \, (\sqrt{t}+\sqrt{t+\xi})^2
\,\left( \sqrt{ u(t+\xi)}+\sqrt{t(1+u)} \right) }\;\;\;\;\;\;\nonumber\\
&\approx&-\int_0^1 du \int_0^{\infty} dz 
 \frac{ \sqrt{1+u}}{
 (\sqrt{u}+\sqrt{1+u})^2}
 \frac{1}{\sqrt{z} (1+z) \, (\sqrt{z}+\sqrt{1+z})^2
\,\left( \sqrt{ u(1+z)}+\sqrt{z(1+u)} \right)}\nonumber\\
&&\;\;\;\;\;\;\;\;\;\;\;\;\;\;\;\;\;\;\;\;\;\;\;\;\;\;\;\;\;\;
\;\;\;\;\;\;\;\;\;\;\;\;\;\;\;\;\;\;\;\;\;\;\;\;\;\;\;\;\;\;
\;\;\;\;\;\;\;\;\;\;\;\;\;\;\;\;\;\;\;\;\;\;\;\;\;\;\;\;\;\;
\;\;\;\;\;\;\;\;\;\;\;\;\;\;\;\;\;\;\;\;\;\;\;\;\;\;\;\;\;\;
+O(\xi)\nonumber \\
&\approx&\argsinh 1\,(\ln 2-3)+\sqrt{2} \,( 1-3\ln 2)-1+6\ln 2+O(\xi)
\label{B:I1d}
\end{eqnarray}

Thus we have, for $\xi \rightarrow 0$
\begin{eqnarray}
\label{B:I1expand}
f_1(\xi) \approx  \ln \xi  \left(1-\sqrt{2}-\argsinh 1 \right)\,  
+\left( 1-\sqrt{2} +\argsinh 1 \: (4 \ln 2-5)+\ln 2\, (4 \sqrt{2}-2)
  \right)+O\left(\xi \ln\xi\right)
\end{eqnarray}

The same method of expansion aplied to $f_2$ gives
\begin{equation}
\label{B:I2split}
f_2(\xi)=f_3(\xi)+f_4(\xi)
\end{equation}
with
\begin{eqnarray}
f_3(\xi)&=&\int_{\xi}^1 \frac{dy}{y \, \sqrt{1-y}} \, \int_0^1 dt\,
\frac{\sqrt{ (1-t)}}{ \sqrt{t+\xi} 
 \left(\sqrt{t}+\sqrt{t+\xi} \right)}\nonumber\\
&\approx&\frac{\left( \ln\xi \right)^2}{2}+\ln\xi \,
\left( \frac{3}{2}-3 \ln 2 \right)+ 2 \ln 2\, \left(2 \ln 2-\frac{3}{2}\right)
+O\left(\xi (\ln \xi)^2\right)\label{B:I3}
\end{eqnarray}
and
\begin{eqnarray}
\hspace{ -1cm}
f_4(\xi)&=&-\xi\, \int_{\xi}^1 \frac{dy}{y \, \sqrt{1-y}
\left(\sqrt{y}+\sqrt{y+\xi} \right)} \, \int_0^1 dt\,
\frac{\sqrt{t (1-t)}}{ \sqrt{t+\xi} 
 \left(\sqrt{t}+\sqrt{t+\xi} \right)\left(\sqrt{t (y+\xi)}+\sqrt{y (t+\xi)} 
\right) 
}\nonumber\\
&\approx&\left(-\frac{3}{2}-\ln 2+\sqrt{2}+\argsinh 1 \right) \, \ln\xi
\nonumber\\
&& \;\;\;\;\;\;\;\;\;\;\;+
\left( (5-4 \ln 2)\,  \argsinh 1+4 (\ln 2)^2+\ln 2\, (1-4 \sqrt{2})
-2+ \sqrt{2}\right) +O(\xi (\ln \xi)^2) \label{B:I4}
\end{eqnarray}
(for the expansion of $f_4$, the same method of splitting
has again been applied)
Hence
\begin{eqnarray}
f_2(\xi)&=&f_3(\xi)+f_4(\xi)\approx
\frac{\left( \ln\xi \right)^2}{2}+\ln\xi \,
\left( \sqrt{2}+\argsinh 1 -4 \ln 2 \right)\nonumber\\
&& \;\;\;\;\;\;\;\;\;\;\;+
\left( (5-4 \ln 2)\,  \argsinh 1+8 (\ln 2)^2+\ln 2\, (-2-4 \sqrt{2})
-2+ \sqrt{2}
\right) +O(\xi (\ln \xi)^2) \label{B:I2expand}
\end{eqnarray}
and, as $I(\xi)=-2 +\xi \: \big( f_1(\xi)+f_2(\xi)\big) $, we get
(with $X=\ln 2-\frac{\ln \xi}{4}$)
\begin{eqnarray}
I(\xi)
&=&-2 +\xi \:   \Big[\frac{\left( \ln\xi \right)^2}{2}+\ln\xi\, \left(
1-4 \ln 2   \right)
+\left(
8 (\ln 2)^2-4 \ln 2\,
-1
\right)
\Big]
+O(\xi^2\, (\ln \xi)^2)\;\;\;\;\;\;\;\;\nonumber\\
&=&-2 +\xi \:   \Big[
8 X^2-4 X-1
\Big]
+O(\xi^2\, X^2)\label{B:Iexpand}
\end{eqnarray}


\begin{thebibliography}{99}
\bibitem{gennes}  P.~G. de Gennes, \textit{J. Chem. Phys.},
   \textbf{48}, 2257 (1968).

\bibitem{fisher-huse} D.~A. Huse and M.~E. Fisher, \textit{Phys. Rev. B}
\textbf{29}, 239 (1984).

\bibitem{fisher} M.~E. Fisher, \textit{J. Stat. Phys.},
    \textbf{34}, 667 (1984).

\bibitem{Essam} J.~W. Essam and A.~J. Guttmann, \textit{Phys. Rev. E},
    \textbf{52}, 5849 (1995).

\bibitem{Einstein} For a brief review see T.~L. Einstein, {\it
Ann. Henri Poincar\'e 4}, Suppl. 2, S811-S824 
(2003); also available in arXiv:cond-mat/0306347.

\bibitem{Richards} H.~L. Richards and T.~L. Einstein, \textit{Phys. Rev. E} ,
     \textbf{72},  016124 (2005).

\bibitem{Ferrari} P. Ferrari and  M. Praehofer, \textit{Markov
      Processes Relat. Fields}, \textbf{12}, 203 (2006).

\bibitem{JohanssonRMT} K. Johansson, \textit{Probab. Theory Rel.},
    \textbf{123}, 225 (2002).

\bibitem{Katori3} M. Katori and H. Tanemura,
  \textit{Phys. Rev. E},
     \textbf{66}, 011105 (2002).

\bibitem{Katori2} M. Katori, H. Tanemura, T. Nagao and N. Komatsuda
  \textit{Phys. Rev. E},
     \textbf{68}, 021112 (2003).

\bibitem{Katori1} M. Katori and H. Tanemura, \textit{J. Math. Phys.},
     \textbf{45}, 3058 (2004).

\bibitem{T-W} C.~A. Tracy and H. Widom, \textit{The Annals of Applied Prob.},
 \textbf{17}, 953 (2007).


\bibitem{Schehr}
 G. Schehr, S.~N. Majumdar, A. Comtet, J. Randon-Furling, \textit{Phys. Rev.
   Lett.}, 
\textbf{101}, 150601 (2008).

\bibitem{N Kobayashi} N. Kobayashi, M. Izumi and M. Katori,
 {\em Phys. Rev. E},  {\bf 78}, 051102 (2008).

\bibitem{Wishart} J. Wishart, {\it Biometrica}, {\bf 20}, 32 (1928).


\bibitem{Wilks} S.S. Wilks, {\em Mathematical Statistics} (John Wiley \& Sons, New York, 1962).


\bibitem{Fukunaga} K. Fukunaga, {\em Introduction to Statistical
    Pattern Recognition} (Elsevier, New York, 1990).

\bibitem{Smith} L.I. Smith, ``A tutorial on Principal Components
  Analysis'' (2002).

\bibitem{arrays1} N. Holter {\it et al.}, {\it Proc. Nat. Acad. Sci. USA}, {\bf
    97}, 8409 (2000).

\bibitem{arrays2} O. Alter {\it et al.}, {\it Proc. Nat. Acad. Sci. USA}, {\bf
    97}, 10101 (2000).

\bibitem{Cavalli}
L.-L. Cavalli-Sforza, P. Menozzi and A. Piazza, 
``The History and Geography of Human Genes'', Princeton Univ. Press (1994).

\bibitem{Patterson} N. Patterson, A. L. Preis and D. Reich, {\it PLoS
Genetics}, {\bf 2}, 2074 (2006).

\bibitem{genetics} J. Novembre and M. Stephens, {\it  Nature Genetics}, {\bf
    40}, 646 (2008).

\bibitem{BP} J.-P. Bouchaud and M. Potters, {\em Theory of Financial
    Risks} (Cambridge University Press, Cambridge, 2001).

\bibitem{Burda} Z. Burda and J. Jurkiewicz, {\it Physica A}, {\bf  344}, 
67 (2004);
Z. Burda, J. Jurkiewicz and B. Waclaw, {\it Acta Physica Polonica}, {\bf
B 36}, 2641 (2005) and references therein.

\bibitem{Preisendorfer} R.W. Preisendorfer, {\em Principal Component
    Analysis in Meteorology and Oceanography} (Elsevier, New York,
  1988)./


\bibitem{James} A.T. James, {\it Ann. Math. Statistics}, {\bf 35}, 475 (1964).

\bibitem{Altland} A. Altland and M. R. Zirnbauer, {\it Phys. Rev. Lett.},
{\bf 76}, 3420 (1996).

\bibitem{Katori4} M. Katori and H. Tanemura, {\it Probab. Theory Relat.
    Fields},
{\bf 138}, 113 (2007).

\bibitem{Forgacs} G. Forgacs, R. Lipowsky, and Th. M. Nieuwenhuizen,
in {\em Phase
Transitions and Critical Phenomena} ed. by C. Domb 
and J.L. Lebowitz (Academic Press,
London, 1991), vol 14, 136 (1991).

\bibitem{Bray} A.J. Bray and K. Winkler, \textit{ J. Phys. A: Math. Gen.}
\textbf{37}, 5493 (2004).

\bibitem{Johansson} K. Johansson, \textit{Comm. Math. Phys.}, {\bf 209}, 
437 (2000).

\bibitem{Johnstone} I.~M. Johnstone, \textit{Ann. Statist.}, {\bf 29},
 295 (2001).

\bibitem{T-W mean} C.~A. Tracy and H. Widom, \textit{Comm. Math. Phys.},
\textbf{159}, 151 (1994); {\bf 177}, 727 (1996).

\bibitem{Mehta} M.~L. Mehta, \textit{Random matrices} (Academic Press, 1991).

\bibitem{Moser} J. Moser, {\it Adv. Math.}, {\bf 16}, 1 (1975).

\bibitem{OP} M.A. Olshanetsky and A.M. Perelomov, {\it Phys. Rep.},
 {\bf 94}, 313 (1983).

\bibitem{YKY} T. Yamamoto, N. Kawakami, and S. Yang, {\it J. Phys.
 A: Math. Gen.}, {\bf 29}, 317 (1996).

\bibitem{marcenko} V.~A. Mar\u cenko, L. A. Pastur, \textit{Math. USSR-Sb},
  \textbf{1}, 457 (1967).

\bibitem{vivo} P. Vivo, S. N. Majumdar, O. Bohigas,
\textit{J. Phys. A: Math. Theor. },
\textbf{40}, 4317-4337  (2007).

 \bibitem{Dean} D.~S. Dean and S.~N. Majumdar, {\it Phys. Rev. Lett.},
 {\bf 97}, 160201 (2006);
\textit{Phys. Rev. E}, \textbf{77}, 041108 (2008).

\bibitem{vergassola}  S.~N. Majumdar and M. Vergassola,
\textit{Phys. Rev. Lett.}, \textbf{102}, 060601 (2009).

\bibitem{Forrester} P. J. Forrester, {\it Nucl. Phys. B}, {\bf 402}, 709
  (1993).

\bibitem{TWbis} C.A. Tracy and H. Widom, {\it Comm. Math. Phys.},
{\bf 161}, 289 (1994).

\bibitem{edelman} A.~Edelman,
 \textit{ J. Matrix Anal. and Appl.} ,\textbf{9}, 543 (1988). 

\bibitem{majumdar-quant} S. N. Majumdar, O. Bohigas, A. Lakshminarayan,
\textit{        J. Stat. Phys.}, \textbf{131},  33 (2008).

\bibitem{facchi} P. Facchi, U. Marzolino, G. Parisi, S. Pascazio, and A. Scardicchio,
\textit{Phys. Rev. Lett.}, \textbf{ 101}, 050502 (2008).

\bibitem{shotnoise} P. Vivo, S. N. Majumdar, O. Bohigas, \textit{Phys. Rev. Lett.},
\textbf{101}, 216809 (2008).

\bibitem{tricomi} F. G. Tricomi, \textit{Integral Equations},
Pure Appl. Math. V, Interscience, London (1957).


    

\end{thebibliography}
\end{document}